\documentclass[12pt,preprint,a4paper]{aastex}

\usepackage{amsmath}

\newcommand{\kms}{km s$^{-1}$}

\newcommand{\luergarc}{erg s$^{-1}$ cm$^{-2}$ arcsec$^{-2}$}
\newcommand{\Msyr}{$M_\odot$ yr$^{-1}$}
\newcommand{\ncm}{cm$^{-3}$}

\newcommand{\um}{$\mu$m}

\newcommand{\Ha}{H${\alpha}$}
\newcommand{\Htwo}{H$_2$}
\newcommand{\HtwolineK}{H$_2$ $\upsilon=1\rightarrow0$ $S$(1)}

\newcommand{\astN}[1]{N {\small #1}}

\newcommand{\astFe}[1]{Fe {\small #1}}

\newcommand{\EBV}{$E(B-V)$}

\newcommand{\Av}{$A_V$}

\newcommand{\Ms}{$M_\odot$}

\newcommand{\Mout}{$\dot{M}_{out}$}

\newcommand{\Maccenv}{$\dot{M}_{acc,env}$}
\newcommand{\Maccdis}{$\dot{M}_{acc,disk}$}

\newcommand{\aatiris}{\textit{AAT}-IRIS2}

\newcommand{\hst}{\textit{Hubble Space Telescope}}
\newcommand{\hstacs}{\textit{HST}-ACS}

\newcommand{\spitzer}{\textit{Spitzer}}

\shorttitle{[\astFe{II}] Jets and Outflows in the Carina Nebula}
\shortauthors{Shinn et al.}

\begin{document}

\title{[\astFe{II}] 1.64 \um{} Features of Jets and Outflows from Young Stellar Objects in the Carina Nebula}

\author{Jong-Ho Shinn\altaffilmark{1}, Tae-Soo Pyo\altaffilmark{2}, Jae-Joon Lee\altaffilmark{1}, Ho-Gyu Lee\altaffilmark{3}, Hyun-Jeong Kim\altaffilmark{4}, Bon-Chul Koo\altaffilmark{4}, Hwankyung Sung\altaffilmark{5}, Moo Young Chun\altaffilmark{1}, A.-Ran Lyo\altaffilmark{1}, Dae-Sik Moon\altaffilmark{1,6}, Jaemann Kyeong\altaffilmark{1}, Byeong-Gon Park\altaffilmark{1}, Hyeonoh Hur\altaffilmark{5}, Yong-Hyun Lee\altaffilmark{4}}

\email{jhshinn@kasi.re.kr}
\altaffiltext{1}{Korea Astronomy and Space Science Institute, 776 Daeduk-daero, Yuseong-gu, Daejeon, 305-348, the Republic of Korea}
\altaffiltext{2}{Subaru Telescope, National Astronomical Observatory of Japan, 650 North A`oh\={o}k\={u} Place, Hilo, HI 96720, U.S.A.}
\altaffiltext{3}{Department of Astronomy, Graduate School of Science, the University of Tokyo, 7-3-1 Hongo, Bunkyo-ku, Tokyo 113-0033, Japan}
\altaffiltext{4}{Dept. of Physics and Astronomy, Seoul National University, 599 Gwanangno, Gwanak-gu, Seoul, 151-747, the Republic of Korea}
\altaffiltext{5}{Department of Astronomy and Space Science, Sejong University, 98 Kunja-dong, Kwangjin-gu, Seoul, 143-747, the Republic of Korea}
\altaffiltext{6}{Department of Astronomy and Astrophysics, University of Toronto, Toronto, ON M5S 3H4, Canada}

\begin{abstract}
We present [\astFe{II}] 1.64 \um{} imaging observations for jets and outflows from young stellar objects (YSOs) over the northern part ($\sim24'\times45'$) of the Carina Nebula, a massive star forming region.
The observations were performed with IRIS2 of Anglo-Australian Telescope and the seeing was $\sim1.5''\pm0.5''$.
Eleven jet and outflow features are detected at eight different regions, and are named as Ionized Fe Objects (IFOs).
One Herbig-Haro object candidate missed in \hst{} \Ha{} observations is newly identified as HHc-16, referring our [\astFe{II}] images.
IFOs have knotty or longish shapes, and the detection rate of IFOs against previously identified YSOs is 1.4 \%, which should be treated as a lower limit.
Four IFOs show an anti-correlated peak intensities in [\astFe{II}] and \Ha, where the ratio $I$([\astFe{II}])/$I$(\Ha{}) is higher for longish IFOs than for knotty IFOs.
We estimate the outflow mass loss rate from the [\astFe{II}] flux, using two different methods.
The jet-driving objects are identified for three IFOs (IFO-2, -4, and -7), for which we study the relations between the outflow mass loss rate and the YSO physical parameters from the radiative transfer model fitting.
The ratios of the outflow mass loss rate over the disk accretion rate are consistent for IFO-4 and -7 with the previously reported values ($10^{-2}-10^{+1}$), while it is higher for IFO-2.
This excess may be from the underestimation of the disk accretion rate.
The jet-driving objects are likely to be low- or intermediate-mass stars.
Other YSO physical parameters, such as luminosity and age, show reasonable relations or trends. 
\end{abstract}


\section{Introduction \label{intro}}
Star formations near massive stars ($\ga8 M_{\odot}$) are peculiar in many aspects.
For instance, strong ultraviolet radiation and stellar winds of massive stars excavate the ambient clouds, which can be the birthplace of new stars, and the ambient clouds can be contaminated with newly-synthesized elements through the supernova explosions of massive stars.
Besides, ultraviolet radiation of massive stars strips away the circumstellar envelope of young stellar objects (YSOs) under formation, and irradiates their jets \citep{Reipurth(1998)Nature_396_343,Reipurth(2001)ARA&A_39_403}, whose full-body shapes are visible.
Star formations near massive stars are even intriguing, since our solar system may emerge from a massive star forming region rather than an isolated low-mass ($\la8 M_{\odot}$) star forming region, as indicated by the contamination of early solar system by short-lived radionuclides \citep{Tachibana(2003)ApJ_588_L41,Hester(2004)Science_304_1116,Mostefaoui(2005)ApJ_625_271,Tachibana(2006)ApJ_639_L87}. 

The Carina Nebula (NGC 3372; $l=287.61,\,b=0.85$) is one of such star forming regions that contains numerous massive stars, locating at $2.2-2.9$ kpc; the distance varies depending on the estimation methods \citep[e.g.][]{Allen(1993)PASAu_10_338,Walborn(1995)inproc,Vazquez(1996)A&AS_116_75,Meaburn(1999)inproc,Davidson(2001)AJ_121_1569,Tapia(2003)MNRAS_339_44,Smith(2006)ApJ_644_1151,Hur(2012)AJ_143_41}.
It contains tens of O-type stars, Wolf-Rayet stars, and star clusters such as Tr 14, Tr 15, and Tr 16 \citep{Smith(2006)MNRAS_367_763}; Tr 16 includes the famous luminous blue variable, $\eta$ Carinae \citep{Davidson(1997)ARA&A_35_1}.
The global on-going star formation in the Carina Nebula has been advocated by several studies \citep[e.g.][]{Megeath(1996)A&A_305_296,Smith(2000)ApJ_532_L145,Rathborne(2004)A&A_418_563,Smith(2010)MNRAS_406_952,Povich(2011)ApJS_194_14}. 
Such an on-going star formation makes one expect the existence of jets and outflows from YSOs through accretion processes, and indeed several Herbig-Haro (HH) objects were observed over the Carina Nebula.
HH 666 (the Axis of Evil) is the first-discovered HH object \citep{Smith(2004)AJ_127_2793} and additional tens of HH objects and HH candidates were reported from \hst{} (\textit{HST}) observations \citep{Smith(2010)MNRAS_405_1153}.

Jets and outflows of YSOs are important in star formation, because they remove the angular momentum from the disk material and enable the mass accretion onto the central object \citep{Shang(2007)2007prpl.conf__261,Pudritz(2007)2007prpl.conf__277}.
Therefore, the outflow phenomenon is closely related to the accretion process, and its precise characterization helps the quantification of the accretion process.
For example, the direct proportion between the accretion rate and the outflow mass loss rate was revealed from the observations on YSOs \citep{Hartigan(1995)ApJ_452_736,Ellerbroek(2013)A&A_551_A5}, and numerous theoretical studies have been performed to explain such a relation between the two rates \citep[e.g.][]{Pelletier(1992)ApJ_394_117,Wardle(1993)ApJ_410_218,Najita(1994)ApJ_429_808,Tomisaka(1998)ApJ_502_L163,Tomisaka(2002)ApJ_575_306,Hennebelle(2008)A&A_477_9,Seifried(2012)MNRAS_422_347,Sheikhnezami(2012)ApJ_757_65}.
Also, the evolution of jets and outflows gives a picture on how the accretion process evolves \citep[e.g.][]{Bontemps(1996)A&A_311_858,Corcoran(1997)A&A_321_189,Arce(2006)ApJ_646_1070}.

One way to trace the jets and outflows is observing shock-excited \Ha{} feature, i.e.~the HH objects.
However, it is not easy to identify when massive stars exist nearby, because the photoionized \Ha{} features mix with the shock-excited \Ha{} features.
Under this circumstance, [\astFe{II}] $a\,^4D_{7/2}-a\,^4F_{9/2}$ 1.644 \um{} emission line is useful.
It is weak when photoionized than shock-excited \citep{Alonso-Herrero(1997)ApJ_482_747}, and it suffers less extinction than \Ha{} because its wavelength is longer than \Ha{} \citep{Draine(2003)ARA&A_41_241}.
In near-infrared, the [\astFe{II}] emission line, together with the H$_2$ emission line, has been used as an important tracer of jets and outflows in star forming regions \citep[e.g.][]{Reipurth(2000)AJ_120_1449,Davis(2001)MNRAS_326_524,Davis(2003)MNRAS_344_262,Hayashi(2009)ApJ_694_582}.
Here we report [\astFe{II}] 1.64 \um{} imaging observations over the northern part ($\sim24'\times45'$) of the Carina Nebula, performed with IRIS2 of Anglo-Australian Telescope (AAT).
Eleven jet and outflow features were detected at eight different regions.
Notably, one of them has a corresponding \Ha{} feature unlisted in the \textit{HST} \Ha{} observations \citep{Smith(2010)MNRAS_405_1153}.
We estimate the outflow mass loss rates from the [\astFe{II}] fluxes, and discuss them in relation with the physical parameters of the central objects.
The morphology, distribution, and detection rate of [\astFe{II}] features are also discussed.

\section{Observations and Data Reduction \label{obs-red}}
The imaging observations were performed with IRIS2 \citep{Tinney(2004)inproc} on the 3.9 m Anglo-Australian Telescope, during 2011 February and March.
IRIS2 is a near-infrared ($0.9-2.5$ \um) imager and spectrograph, based on a $1024\times1024$  Rockwell HAWAII-1 HgCdTe infrared detector.
It provides wide-field ($7.7'\times7.7'$) imaging capabilities at $0.4486''$ per pixel sampling.
Its zeropoint magnitude at H band is 22.79\footnote{This is from the IRIS2 webpage, http://goo.gl/XWTb9}.
The full-width-at-half-maximum seeing ranged from $1.1''$ to $2.0''$ and its median was $1.5''$.

\begin{figure}
\center{
\includegraphics[scale=0.7]{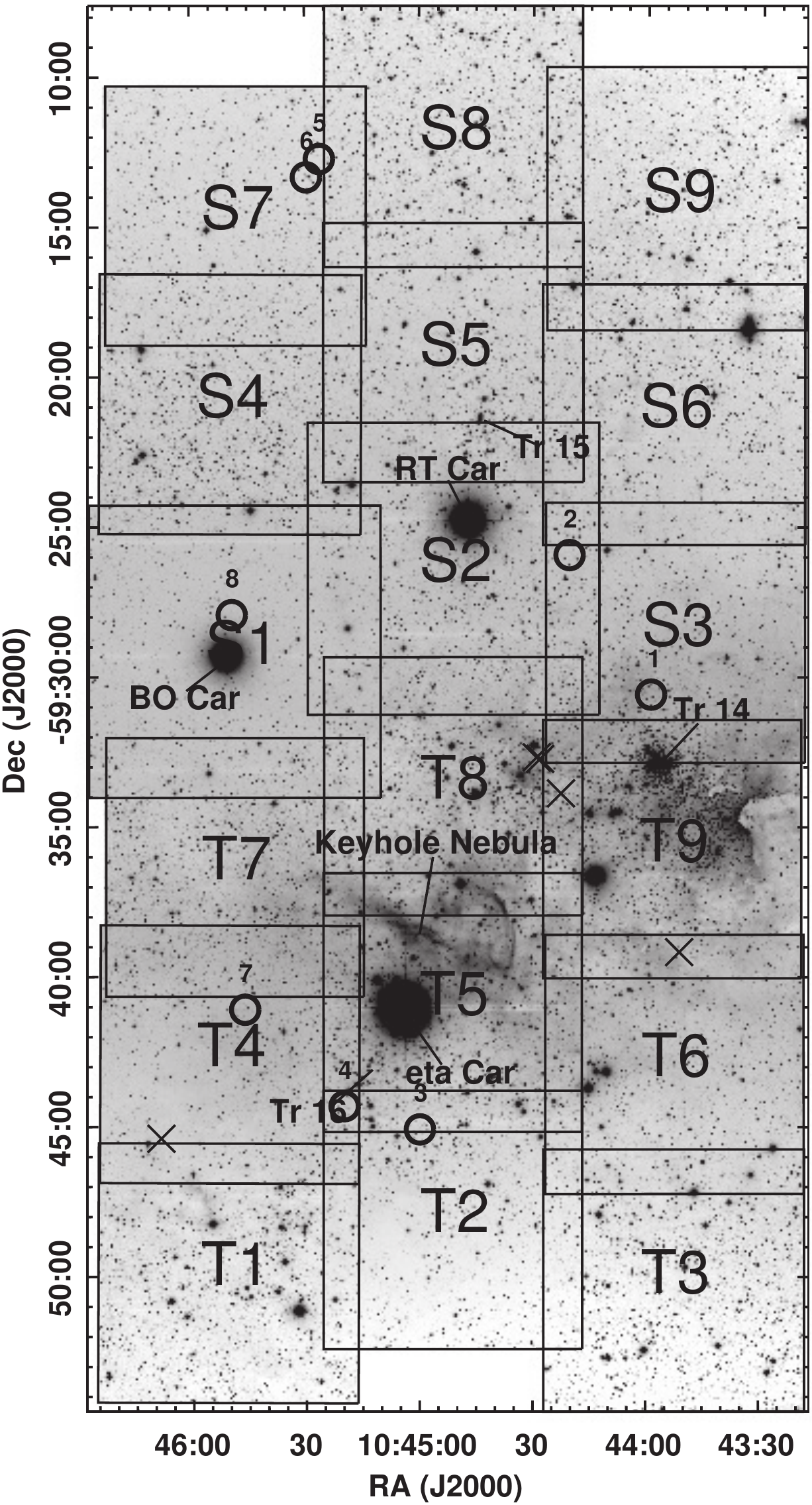}
}
\caption{The [\astFe{II}] 1.64 \um{} filter image observed with \aatiris. The mosaic image is made from the 18 smaller images. The box indicates each pointing, and its region name is shown (T1-T9, S1-S9). The regions are overlapped with each other. The black circles are the IFO locations with numbering (cf.~Table \ref{tbl-flux}). The black ``$\times$'' indicate the location of five MHOs. The names of some bright features are indicated with lines. The coordinates are RA and Dec in J2000. \label{fig-obs}}
\end{figure}

The observed region is shown in Figure \ref{fig-obs} and the detailed observation information is given in Table \ref{tbl-obs}.
The $\sim24'\times45'$ region was covered by 18 different pointings, which are named as T1-T9 and S1-S9.
We observed these 18 regions in pairs (target, sky) with two filters: H ($\lambda_c=1.633$ \um, $\Delta\lambda_{equiv}=257$ nm) and [Fe II] ($\lambda_c=1.633$ \um, $\Delta\lambda_{equiv}=23.4$ nm), where $\lambda_c$ and $\Delta\lambda_{equiv}$ are the center wavelength and the equivalent width of the filter, respectively.
The target-sky pairs are as follows: (T1,S1), (T2,S2), (T4,S4), (T5,S5), (T6,T3), (T7,S7), (T8,S8), (T9,S6), and (S3,S9).
When the target regions are dominated by nebulosities, and thus generating self-flat and self-sky images is difficult, we observed the target and sky positions alternately; this method was applied only to the [\astFe{II}] filter, since the exposure time for the H filter is short enough to ignore the sky variation.
This alternate exposure method was used for all the pairs, except (T1,S1) and (T2,S2).

We used 9 point dithering for all the observations, and the dithering width was $30''$, except for S1 and S2; a $1'$ dithering was used for S1 and S2, since two very bright sources are within the field-of-view (see Fig.~\ref{fig-obs}).
For the alternate exposure, the dithering starts and ends at the sky position like (sky-target-sky-...-sky).
Therefore, the target and sky positions were exposed 9 and 10 times, respectively.
The exposure times were basically 90 s and 1800-2000 s for H and [\astFe{II}], respectively (Table \ref{tbl-obs}).
Depending on the brightness of the stars within the field-of-view, we adjusted the cycle and period of the exposure.

The data reduction was carried out employing ORAC-DR \citep{Cavanagh(2008)AN_329_295}, which is bundled with the Starlink Software Collection \citep{Jenness(2009)inproc}.
The raw images for the individual regions were separately processed.
The sky and flat images were generated from the raw image of the sky-positions for all the pairs, except for two pairs (T1,S1) and (T2,S2).
The self-sky and self-flat images were generated for T1 and T2, while only the self-sky image was generated for S1 and S2.
We used the flat image of S7 for S1 and S2, since very bright sources within the field hinder the self-flat generation (cf.~Fig.~\ref{fig-obs}) and the S7 region was observed just before observing S1 and S2.
The sky images were generated by median-averaging the dithered raw images with masking bright point sources.
After subtracting the sky image from each raw image, the final median-coadded image was obtained through several processes, such as bad-pixel masking, flat-fielding, distortion correction, feature detection and matching between object frames, and resampling.
Finally, the astrometric information was given to the coadded image by correlating the positions of point sources in the image to the 2MASS point source catalog \citep{Skrutskie(2006)AJ_131_1163}.

The obtained coadded [\astFe{II}] image was flux-calibrated at each region.
We compared the countrate of the 12-23 field point sources with the corresponding H magnitude ($\sim12-15$ mag) from the 2MASS catalog, compensating the smaller equivalent width of the [\astFe{II}] filter.
The linear Pearson correlation coefficient between the two quantities were 0.998-0.999.
The calibration errors, which is dominated by 2MASS photometric errors, were 8-20 \%.
We included them in the error estimation.

\section{Analysis and Results \label{ana-res}}
\subsection{[\astFe{II}] Detection of Jets and Outflow Features \label{ana-res-det}}
In order to detect jet and outflow features in the [\astFe{II}] images, we made continuum-subtracted [\astFe{II}] images using the H images.
Both [\astFe{II}] and H images were resampled onto a common grid.
Then, a background image, made by median-smoothing with a $\sim3'\times3'$ box, was subtracted from the resampled image for the removal of any large-scale background variation.
From these images, we made the continuum-subtracted [\astFe{II}] image by subtracting the scaled H images from the [\astFe{II}] images.
The scaling factor was found to make the pixel value distribution in the resultant image to be symmetric around the median pixel values, referring the exposure period and the equivalent width of filters.

We found 11 [\astFe{II}] features at eight different areas, and named them as Ionized Fe Object (IFO) with numbering.
Their positions are indicated in Figure \ref{fig-obs} and listed in Table \ref{tbl-flux}, and their images are shown in Figure \ref{fig-jet}.
In Figure \ref{fig-jet}, we mark the YSOs identified by \cite{Povich(2011)ApJS_194_14} as diamonds with colors indicating the evolutionary stages of YSOs.
The IFOs generally show knotty or longish shapes.
We give a brief description on the individual IFOs below.
Four of 11 IFOs are related to four HH objects---HH 900, HH 902, HH 1013 NE2, and HH 1014 (Table \ref{tbl-flux})---discovered by \cite{Smith(2010)MNRAS_405_1153} using the Advanced Camera for Surveys (ACS) of \textit{HST}.
We searched the corresponding \Ha{} features of the other IFOs from the \hstacs{} images, and found one for IFO-3.

\subsubsection{IFO-1}
IFO-1 shows a knotty shape and corresponds to the west tip of HH 902 (Fig.~\ref{fig-jet-a} and \citealt{Smith(2010)MNRAS_405_1153}).
Considering the bow-like appearance seen in the \hstacs{} image and the discussion on HH 902 by \cite{Smith(2010)MNRAS_405_1153}, the jet-driving object likely locates in the east of IFO-1.
However, there is no YSO of \cite{Povich(2011)ApJS_194_14} at the expected location.
The filamentary structure of clouds at the location may hinder the identification of YSOs.
We tried to identify any other [\astFe{II}] feature related with HH 902, but it was hard to discern [\astFe{II}] feature from continuum feature because of the low imaging resolution.

\subsubsection{IFO-2 \label{ana-res-det-2}}
IFO-2 falls onto the overlapped region (S2 and S3; Fig.~\ref{fig-obs}), and we have two different images of IFO-2.
We only present the image from S2 region, because it has a better seeing (S2 $\sim1.28''$, S3 $\sim1.50''$).
We did not merge the two images to keep the better seeing.
IFO-2 shows a knotty shape and corresponds to HH 1013 NE2 \citep{Smith(2010)MNRAS_405_1153}.
The jet-driving object identified by \cite{Smith(2010)MNRAS_405_1153} corresponds to the YSO \#490 of \cite{Povich(2011)ApJS_194_14}.
This link between IFO-2 and its driving object enables us to inspect the system in-depth (cf.~section \ref{ana-res-dep}). 
We found no corresponding [\astFe{II}] features for the HH 1013 NE1, SW1, SW2, SW3, and SW4 \citep{Smith(2010)MNRAS_405_1153}.

\subsubsection{IFO-3}
IFO-3 shows a knotty shape (Fig.~\ref{fig-jet-c}).
\cite{Smith(2010)MNRAS_405_1153} identified no corresponding HH object in their \hstacs{} observations; however, we identify an \Ha{} feature not catalogued by \cite{Smith(2010)MNRAS_405_1153} that coincides with IFO-3  (see Fig.~\ref{fig-jet-c}).
Its surface brightness is $\sim4.8\times10^{-15}$ \luergarc, which is similar to other HH objects' reported by \cite{Smith(2010)MNRAS_405_1153}.
This \Ha{} feature can be an HH object and we thus named it as HHc-16 following \cite{Smith(2010)MNRAS_405_1153}.
The appearance of HHc-16 is not easy to discern from those of other photo-ionized features, and this fact highlights the usefulness of [\astFe{II}] 1.64 \um{} line for confirming shock-excited \Ha{} features under strong ultraviolet radiation environments (cf.~section \ref{intro}).

There is no evident infrared point source that seems to relate with IFO-3.
The shape seen in the \Ha{} image (Fig.~\ref{fig-jet-c}) is filamentary and elongated along the northeast-southwest direction.
If IFO-3 is an outflow feature of a YSO and the \Ha{} features represent multiple shell-like shock feature, the YSO likely locates in the northwestern or southeastern side of IFO-3.
In this case, the YSOs \#718, \#725, and \#754 of \cite{Povich(2011)ApJS_194_14} are the candidates.
On the other hand, IFO-3 might be a sharp-edged bow shock feature propagating southwestward, although there is no YSO candidate of \cite{Povich(2011)ApJS_194_14} northeastward from IFO-3 within $\sim0.8$ pc (cf.~Fig.\ref{fig-jet-c}).
In any case, it is hard to pin down only based on the imaging data.
Kinematic studies are required for the identification of IFO-3 as an outflow feature and of its driving object.

\subsubsection{IFO-4 \label{ana-res-det-4}}
IFO-4 falls onto the overlapped region of T2, T4, and T5 (Fig.~\ref{fig-obs}), but it is not detected in T5 due to the poor seeing.
We only present the image from T4 region, because it has a better seeing (T2 $\sim1.58''$, T4 $\sim1.29''$).
We did not merge the two images to keep the better seeing. 
IFO-4 shows a longish shape in the northeast-southwest direction and corresponds to HH 900 \citep{Smith(2010)MNRAS_405_1153}.
The jet-driving object identified by \cite{Smith(2010)MNRAS_405_1153} corresponds to the YSO \#842 of \cite{Povich(2011)ApJS_194_14}.
This link between IFO-4 and its driving object enables us to inspect the system in-depth (cf.~section \ref{ana-res-dep}). 
We found no corresponding [\astFe{II}] feature for HH 900 NE, while there is a probable one for HH 900 SW \citep[cf.][]{Smith(2010)MNRAS_405_1153}.
However, HH 900 SW seems to be mixed with a point source (Fig.~\ref{fig-jet-d}) and hence we did not perform the flux measurement for it (cf.~section \ref{ana-res-flux}).

\subsubsection{IFO-5}
IFO-5 shows a longish shape in the northwest-southeast direction (Fig.~\ref{fig-jet-e}) and was not covered by the \hstacs{} \Ha{} observations \citep{Smith(2010)MNRAS_405_1153}.
Its driving object is probably the closest northern point source, because IFO-5 seems to emerge from the source (cf.~the arrow in Fig.~\ref{fig-jet-e}).
This point source has no counterpart in the 2MASS catalog \citep{Skrutskie(2006)AJ_131_1163} and the \spitzer{} IRAC images (the Vela-Carina survey, \spitzer{} Proposal ID 40791, \citealt{Zasowski(2009)ApJ_707_510}), except a marginal one in the IRAC 8.0 \um{} image.
This non-detection is caused by the shallower depth of the images.
The 2MASS and IRAC images have a similar point source sensitivity  \citep{Benjamin(2003)PASP_115_953}, where [3.6] $\sim15.5$ mag \citep{Zasowski(2009)ApJ_707_510}.
Note that this candidate was not analyzed by \cite{Povich(2011)ApJS_194_14}, since it has a photometric information only in the 8.0 \um{} IRAC band, among 2MASS and \spitzer{} IRAC bands.
If IFO-5 is not emerging from the candidate mentioned above, the jet-driving object can be other YSOs.
Considering its bow-like shape, the jet-driving object may exist northeastward from IFO-5.
Other nearby YSOs identified by \cite{Povich(2011)ApJS_194_14} such as \#884 or \#879 are another candidate in general (cf.~Fig.~\ref{fig-jet-e}).
Higher spatial-resolution images would significantly contribute to the identification of the jet-driving object.

\subsubsection{IFO-6}
IFO-6a shows a longish shape in the east-west direction, while IFO-6b and IFO-6c show a knotty shape (Fig.~\ref{fig-jet-f}).
IFO-6 was not covered by the \hstacs{} \Ha{} observations \citep{Smith(2010)MNRAS_405_1153}.
There are three YSOs of \cite{Povich(2011)ApJS_194_14} around IFO-6: \#879, \#884, and \#889.
Considering the direction that three features of IFO-6 distribute, the jet-driving object is likely either \#884 or \#889, or both.
Kinematic studies of IFO-6 with a better imaging resolution is required to pin down their jet-driving objects.

\subsubsection{IFO-7}
IFO-7 shows an evident bipolar [\astFe{II}] feature in the east-west direction (Fig.~\ref{fig-jet-g}), and corresponds to HH 1014 \citep{Smith(2010)MNRAS_405_1153}.
\cite{Smith(2010)MNRAS_405_1153} described HH 1014 as an one-sided, westward jet.
However, [\astFe{II}] features manifest the eastward counter-jet, which has a probable \Ha{} feature in the \hstacs{} image (Fig.~\ref{fig-jet-g}).
The jet-driving object was expected to be embedded in the cloud \citep{Smith(2010)MNRAS_405_1153}, and indeed there is one YSO (\#984) of \cite{Povich(2011)ApJS_194_14} at the right position.
This link between IFO-7 and its driving object enables us to inspect the system in-depth (cf.~section \ref{ana-res-dep}).

\subsubsection{IFO-8}
IFO-8 shows a knotty shape (Fig.~\ref{fig-jet-h}) and was not covered by the \hstacs{} \Ha{} observations \citep{Smith(2010)MNRAS_405_1153}.
There is no evident infrared point source that seems to relate with IFO-8.
There are three YSOs of \cite{Povich(2011)ApJS_194_14} around IFO-8: \#1007, \#1010, and \#1032.
However, it is hard to pin down the jet-driving object only based on the imaging data, if IFO-8 is an outflow feature of a YSO.
Kinematic studies are required for the identification of IFO-8 as an outflow feature and of the driving object.

\subsection{[\astFe{II}] Flux Measurements and Outflow Mass Loss Rate Estimation \label{ana-res-flux}}
We measured the [\astFe{II}] flux of IFOs.
The source areas were carefully determined to include all the plausible sources from the [\astFe{II}] images, and the corresponding background areas were chosen nearby.
The source and background areas are seen in Figure \ref{fig-jet}.
The measured [\astFe{II}] fluxes are then extinction-corrected.
We adopt a line-of sight extinction of $A_V=3.5$ for the entire region observed, based on the near-infrared photometric survey of \cite{Preibisch(2011)ApJS_194_10}, and employed the extinction curve of ``Milky Way, $R_V=4.0$'' \citep{Weingartner(2001)ApJ_548_296,Draine(2003)ARA&A_41_241}.
The $R_V=4.0$ curve rather than $R_V=3.1$ was used, because the extinction toward the Carina Nebula is better described by the $R_V=4.0$ curve \citep{Preibisch(2011)ApJS_194_10,Povich(2011)ApJS_194_6}.
$A_{[\astFe{II}]}$ is about 0.58.
The observed and dereddened fluxes are listed in Table \ref{tbl-flux}.
IFO-6a is the strongest and IFO-7b is the weakest, and their fluxes are about tenfold different.
For IFO-2 and IFO-4, the fluxes from two different exposures are coincident within 1-$\sigma$, and we list the averaged values (Table \ref{tbl-flux}).

The outflow mass loss rate was derived from the dereddened [\astFe{II}] flux in two different ways according to the shape of IFOs.
When the shape is knotty, the outflow mass loss rate is estimated on the assumption that the [\astFe{II}] gas is heated by either wind shock or ambient shock \cite[cf.~Fig.~\ref{fig-knot}; section IV of][]{McKee(1987)ApJ_322_275}.
On the other hand, when the shape is longish, the outflow mass loss rate is estimated on the assumption that the [\astFe{II}] gas represents the ejected mass flow; in this case, the [\astFe{II}] gas may be heated by either shocks or radiation, or both.
Table \ref{tbl-out} lists the outflow mass loss rates, which ranges $\sim10^{-7}-10^{-6}$ \Msyr.
For comparison, the outflow mass loss rates estimated from \Ha{} intensity assuming irradiated heating \citep{Smith(2010)MNRAS_405_1153} are also listed.

\subsubsection{Outflow Mass Loss Rate of Knotty IFOs \label{out-knot}}
We estimated the outflow mass loss rate (\Mout), equating the total mechanical luminosity of shock ($L_{mech}$) with the kinetic energy input rate into the shock.
First, $L_{mech}$ is obtained as follows.
\begin{eqnarray}
L_{[\astFe{II}]}=f_{[\astFe{II}]}\times4\pi d^2 \\
L_{mech}=L_{[\astFe{II}]}/(1.5\times10^{-3}) \label{eq-fratio}
\end{eqnarray}
The dereddened [\astFe{II}] flux ($f_{[\astFe{II}]}$) was converted to the [\astFe{II}] luminosity ($L_{[\astFe{II}]}$), using the distance to the Carina Nebula.
The distance varies from 2.2--2.9 kpc, depending on the estimation methods (cf.~section \ref{intro}).
We adopted 2.5 kpc as the distance $d$.
Then, $L_{[\astFe{II}]}$ was converted to $L_{mech}$, using a conversion factor from a model for dissociative radiative shocks \citep{Allen(2008)ApJS_178_20}.
The factor converting the [\astFe{II}] 1.64 \um{} emission line flux to the mechanical input energy flux ($\frac{1}{2}\times\textrm{preshock mass density}\times\textrm{shock velocity}^3$) is expected to be $(0.7-2.0)\times10^{-3}$ for shocks propagating into a neutral atomic gas, where the preshock density ranges $1-10^3$ \ncm{}, the shock velocity ranges $100-300$ \kms{}, and the abundance is solar.
These preshock density and shock velocity are typical for jets and outflows of YSOs \cite[][and references therein]{Reipurth(2001)ARA&A_39_403}.
Note that the conversion factor only varies within a factor of three over the shock velocity range of $100-300$ \kms, which corresponds to a factor of $\sim30$ difference in $L_{mech}$.
It seems that the increase of $L_{mech}$ is compensated by the enhanced hard X-rays and subsequent ionization at the postshock region of a few $10^3$ K \citep{Allen(2008)ApJS_178_20}.
We adopted $1.5\times10^{-3}$ as the nominal conversion factor.

\begin{figure}
\figurenum{3}
\center{
\includegraphics[scale=0.8]{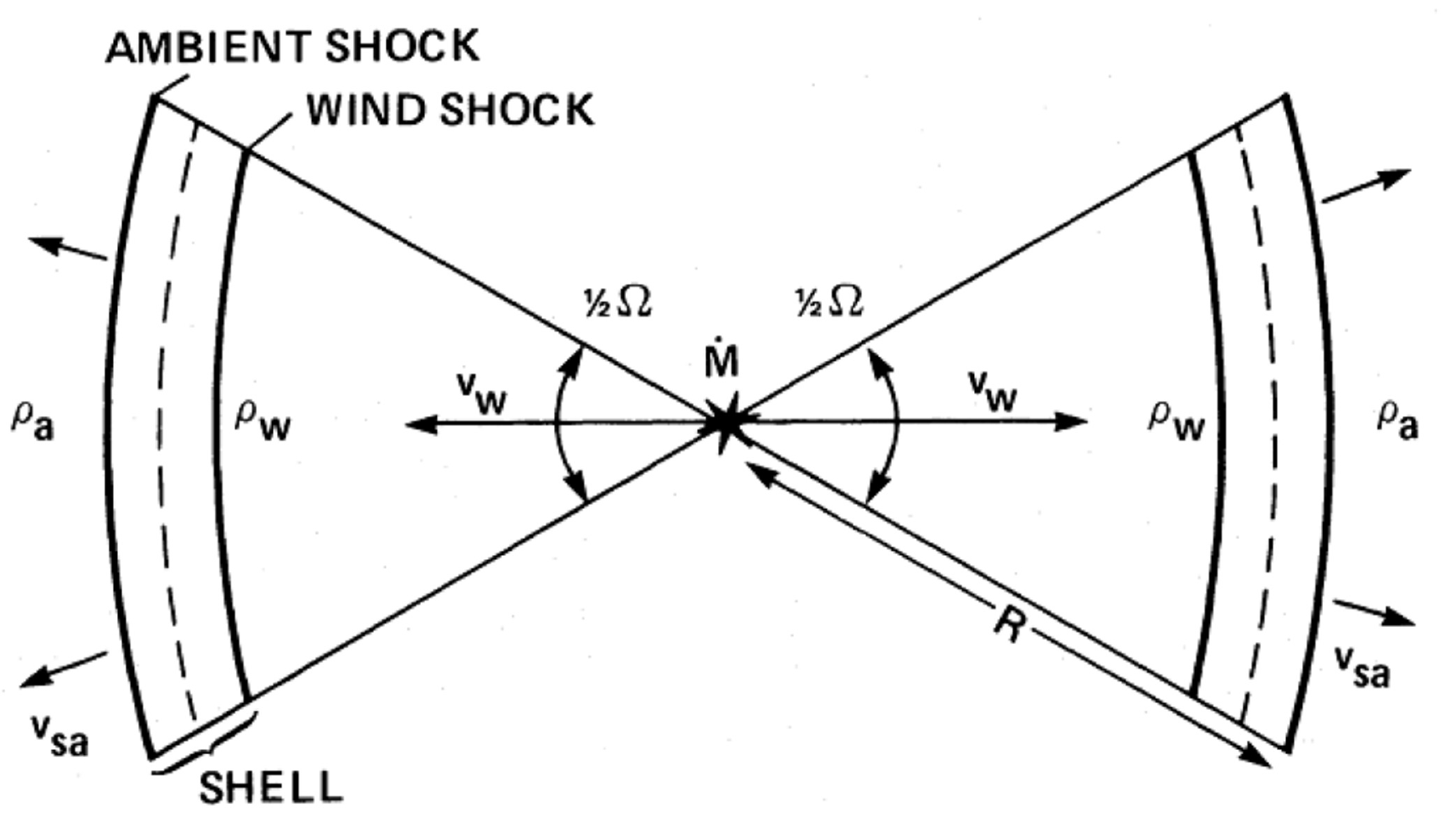}
}
\caption{Schematic diagram for the wind and ambient shocks generated by the outflow of a young stellar object. This figure is taken from \cite{McKee(1987)ApJ_322_275}. The ``shell'' consists of shocked ambient and wind material, which are divided by the dashed curves. \label{fig-knot}}
\end{figure}

Second, the kinetic energy input rate into the shock is differently given at the wind shock and at the ambient shock.
We simplified the shock configuration not to have lateral motions of shocked gas (Fig.~\ref{fig-knot}).
A more realistic configuration may be like the one shown in \cite{Hartigan(1989)ApJ_339_987}.
Following the notation of \cite{McKee(1987)ApJ_322_275}, the kinetic energy input rate into the shock is $\frac{1}{2}\dot{M}_{sw}v^2_{sw}$ at the wind shock, while it is $\frac{1}{2}\dot{M}_{sa}v^2_{sa}$ at the ambient shock.
$\dot{M}_{sw}$ and $\dot{M}_{sa}$ are the mass flow rate into the wind shock and the ambient shock, respectively.
$v_{sw}$ and $v_{sa}$ are the shock velocity at the wind shock and the ambient shock, respectively.
$v_{sw}=v_w-v_{sa}$, where $v_w$ is the wind velocity (see Fig.~\ref{fig-knot}).
Here, \Mout{} is related with $\dot{M}_{sw}$ as below \citep{McKee(1987)ApJ_322_275},
\begin{eqnarray}
\dot{M}_{sw}=(1-v_{sa}/v_w)\dot{M}_{out} \label{eq-msw}
\end{eqnarray}
\Mout{} can also be related with $\dot{M}_{sa}$ through $\dot{M}_{sw}$, using the relation between the driving pressures at both shock sides,
\begin{eqnarray} \label{eq-pres}
\rho^{}_w v^2_{sw}=\frac{2}{3}\rho_a v^2_{sa}
\end{eqnarray}
where $\rho_w$ and $\rho_a$ are the mass density of wind and ambient material, respectively \citep{McKee(1987)ApJ_322_275}.
Eq.~(\ref{eq-pres}) can be transformed to represent the relation of the mass flow rate at both sides as follows.
\begin{eqnarray}
\frac{\rho_w v_{sw}}{\rho_a v_{sa}}=\frac{2}{3}\frac{v_{sa}}{v_{sw}}
\end{eqnarray}
Here, the ratio $\rho_w v_{sw}/\rho_a v_{sa}$ is almost equal to $\dot{M}_{sw}/\dot{M}_{sa}$, since the cross-sectional area is likely comparable if the distance $R$ is sufficiently larger than the thickness of the shocked shell (cf.~Fig.~\ref{fig-knot}).
\begin{eqnarray}
\frac{\dot{M}_{sw}}{\dot{M}_{sa}}=\frac{2}{3}\frac{v_{sa}}{v_{sw}} \label{eq-msa}
\end{eqnarray}

In order to derive \Mout{}, it is needed to determine which shock radiates the observed [\astFe{II}] emission.
We think that both the wind and ambient shocks are J-type and responsible for the [\astFe{II}] as below (for the type of shock, see \citealt{Draine(1993)ARA&A_31_373}).
C-type shock usually shows strong \HtwolineK{} 2.12 \um{} emission \citep[cf.][]{Wilgenbus(2000)A&A_356_1010}.
However, no such emissions related with IFOs were detected even with the 8 m Very Large Telescope \cite[cf.~section \ref{dis-det} and][]{Preibisch(2011)A&A_530_A34}.
Besides, the typical velocity and density of HH objects are suitable for the wind shock to be J-type \citep{Draine(1993)ARA&A_31_373}.
The typical density is $\sim10^2-10^3$ \ncm{} and the typical tangential velocity is $\sim100-500$ \kms{} \cite[][and references therein]{Reipurth(2001)ARA&A_39_403}.
Indeed, a jet within the Carina Nebula (HH 666) shows a velocity of 200 \kms{} \citep{Smith(2004)AJ_127_2793}.
We think the ambient shock can also be J-type unless the ambient medium is too much denser than the wind.

Therefore, we can write the equation for the outflow mass loss rate estimation as below, using eq.~(\ref{eq-msw}) and (\ref{eq-msa}).
\begin{align}
L_{mech} &= \frac{1}{2}\dot{M}_{sw}v^2_{sw} + \frac{1}{2}\dot{M}_{sa}v^2_{sa} = \frac{1}{2}\dot{M}_{sw}v^2_{sw} +\frac{1}{2}\Bigg(\frac{3}{2}\frac{v_{sw}}{v_{sa}}\dot{M}_{sw}\Bigg)v^2_{sa} \nonumber \\
&=\frac{1}{2}\dot{M}_{sw}v^2_{sw}\Bigg(1+\frac{3}{2}\frac{v_{sa}}{v_{sw}}\Bigg)=\frac{1}{2}\Bigg(1-\frac{v_{sa}}{v_{w}}\Bigg)\dot{M}_{out}v^2_{sw}\Bigg(1+\frac{3}{2}\frac{v_{sa}}{v_{sw}}\Bigg) \nonumber \\
&=\frac{1}{2}\Bigg(1-\frac{v_{sa}}{v_{w}}\Bigg)\Bigg(1+\frac{3}{2}\frac{v_{sa}}{v_{sw}}\Bigg)\dot{M}_{out}v^2_{sw} \label{eq-lm-gen}
\end{align}
\cite{McKee(1987)ApJ_322_275} showed that when $R=1.281R_{ch}$, the total energy of the system is minimized for a given kinetic energy of shocked ambient medium; here, $R_{ch}$ is a characteristic length scale.
If we assume this is the case for the observed IFOs (i.e.~$R\sim R_{ch}$), as assumed in \cite{Chernoff(1982)ApJ_259_L97}, then eq.~(\ref{eq-lm-gen}) becomes
\begin{align}
L_{mech}&=\frac{1}{2}\Bigg(1-\frac{1}{3}\Bigg)\Bigg(1+\frac{3}{2}\frac{1}{2}\Bigg)\dot{M}_{out}v^2_{sw}=\frac{7}{12}\dot{M}_{out}v^2_{sw} \nonumber \\
&=\frac{7}{27}\dot{M}_{out}v^2_{w} \label{eq-lm}
\end{align}
where $v_{sw}=\frac{2}{3}v_w$ \citep{McKee(1987)ApJ_322_275}.
Adopting $v_w=200$ \kms{} in eq.~(\ref{eq-lm}), we estimated \Mout{} for the knotty IFOs.

\subsubsection{Outflow Mass Loss Rate of Longish IFOs \label{out-long}}
We assumed that the [\astFe{II}] gas represents a portion of ejected mass flow, and estimated the flow rate of the ejected mass.
Employing the method of \cite{Davis(2003)A&A_397_693}, the outflow mass loss rate ($\dot{M}_{out}$) was derived from the following equations.
\begin{eqnarray}
\frac{f_{[\astFe{II}]}}{\Omega}=\frac{h\nu_{ul}\,N_{Fe^+,u}\,A_{ul}}{4\pi} \\
N_H=\frac{N_{Fe^+,u}}{A_{Fe/H}\,f_{Fe^+}\,f_{Fe^+,u}} \\
M_H=N_H\mu\Omega d^2 \\
\dot{M}_{out}=\frac{M_H}{\tau}=\frac{M_H}{l_{out}/v_{out}} \label{eq-moutir}
\end{eqnarray}
$\Omega$ is the solid angle of the region where the flux is measured (Table \ref{tbl-out}), and $h$ is the Planck constant.
$\nu_{ul}$ and $A_{ul}$ are the frequency and the Einstein coefficient for the [\astFe{II}] 1.64 \um{} transition, respectively.
We used $A_{ul}=0.00465$ s$^{-1}$ \citep{Nussbaumer(1988)A&A_193_327}.
$N_{Fe^+,u}$ and $N_H$ are the column densities of the upper-level and total (atomic and ionic) hydrogen, respectively.
$A_{Fe/H}$ is the abundance of Fe relative to H, and we used $A_{Fe/H}=3.2\times10^{-5}$ \citep{Asplund(2009)ARA&A_47_481}.
$f_{Fe^+}$ and $f_{Fe^+,u}$ are the fraction of Fe$^+$ and of Fe$^+$ in the upper level, respectively.
We adopted $f_{Fe^+}=0.29$ \citep{Hamann(1994)ApJS_93_485} and $f_{Fe^+,u}=0.01$ \citep{Hamann(1994)ApJ_436_292} for the typically observed density $n_e\sim10^4$ \ncm{} and temperature $T\sim10^4$ K.
The adopted $f_{Fe^+}=0.29$ is different from $f_{Fe^+}=0.68$ adopted by \cite{Davis(2003)A&A_397_693}.
$f_{Fe^+}=0.68$ is for $T\sim1.4\times10^4$ K, and we think this temperature is a bit high for the [\astFe{II}] 1.64 \um{} emission, which is a dominant cooling line at $T\simeq5000-10000$ K \citep{Hollenbach(1989)ApJ_342_306}.
$\mu$ is the mean mass per particle, and we used $\mu=1.3\,m_H$ for neutral atomic gas \citep{Hollenbach(1979)ApJS_41_555}, where $m_H$ is the hydrogen mass.
We adopted $d=2.5$ kpc as for the knotty IFOs (cf.~Section \ref{out-knot}).
$l_{out}$ is the outflow length, $v_{out}$ is the outflow velocity that we again assumed 200 \kms.
We set $l_{out}$ as the major axis length of the elliptical regions that used for the flux measurements (Table \ref{tbl-out}).
\Mout{} is generally comparable to those of Class I protostars derived using the same method \citep{Davis(2003)A&A_397_693}, considering the differently adopted $f_{Fe^+}$ mentioned above.

\subsubsection{Caveats for the Outflow Mass Loss Rate Estimation \label{out-cav}}
For the knotty IFOs, \Mout{} is estimated on the assumption that the gas crossing the shock discontinuity is ejected from the jet-driving source \emph{at the same time} and this ejected gas is all shocked without leaving any gas unshocked.
Therefore, \Mout{} is overestimated if the shocked gas is a mixture of materials ejected at different times, while it is underestimated if some ejected gas is not shocked.
\Mout{} may also be underestimated, if the postshock gas does not cool enough because the shock does not sweep enough preshock material to fully develop the postshock structure---i.e. the factor $1.5\times10^{-3}$ in eq.~(\ref{eq-fratio}) can be lower.
The assumption $R\sim R_{ch}$ used for the knotty IFOs is arbitrary; hence, the real \Mout{} can be different from the estimated value depending on the ratio $R/R_{ch}$. 
For the longish IFOs, we ignored the inclination of the outflow with respect to the sky plane.
We assumed that the outflow is on the plane of sky.
Hence, $l_{out}$ in eq. (\ref{eq-moutir}) should be $l_{out}/\cos\theta$, and \Mout{} will be reduced by a factor of $\cos\theta$.
In addition, the extinction correction for the [\astFe{II}] flux can affect the outflow mass loss rate of both the knotty and the longish IFOs.
Early-type stars in the cluster Tr 14 and 16 show \EBV{} $\sim0.3-0.9$ \citep{Hur(2012)AJ_143_41}, while \EBV{} we adopted is $A_V/R_V=3.5/4.0=0.875$.
Therefore, the [\astFe{II}] flux and \Mout{} can be overestimated up to about a factor of 1.4.
The assumed outflow velocity is obviously another caveat for both the knotty and the longish IFOs.

Before closing this section, we here note that the obtained \Mout{} are comparable regardless of the estimation methods, i.e. ``knotty'' or `` longish'' (Table \ref{tbl-out}).
The values fall within a factor of two or three, except for IFO-8.
Even for IFO-8, it falls within about five times the other's mean.
This shows that both estimation methods are comparable, although they are subject to the same errors such as extinction, distance to IFOs, etc.

\subsection{In-depth Inspection on the Three IFOs: IFO-2, -4, and -7 \label{ana-res-dep}}
In section \ref{ana-res-det}, we are able to identify the jet-driving objects for three IFOs: IFO-2, -4, and -7.
More detail inspections for those IFOs in relation with the jet-driving objects are given in this section.
The three jet-driving objects are among the YSOs in the Carina Nebula identified from their infrared excess by \cite{Povich(2011)ApJS_194_14}.
The infrared excess was estimated from 2MASS and \spitzer{} IRAC photometry data ($1-8$ \um), employing the spectral energy distribution (SED) fitting and color criteria.
We first show how we obtain the physical parameters of the jet-driving object from SED fittings, and then inspect individual object.

\subsubsection{Physical Parameters of the Jet-driving YSOs \label{sed-fit}}
The jets and outflows are thought to be launched from the star-disk system \citep{Pudritz(2007)2007prpl.conf__277,Shang(2007)2007prpl.conf__261}.
It is worthwhile to compare the relation between the physical parameters of the star-disk system and the outflows.
\cite{Povich(2011)ApJS_194_14} provided several physical quantities of YSOs in their catalog.
However, some important quantities (e.g.~the disk accretion rate) are not listed.
Therefore, we performed the SED fitting by ourselves to the photometry data provided in the catalog of \cite{Povich(2011)ApJS_194_14}.
We used the SED fitting model of \cite{Robitaille(2007)ApJS_169_328}, and the photometry data are 2MASS, \spitzer-IRAC, and \spitzer-MIPS.
We constrained the distance from 2.2 to 2.9 kpc (cf.~section \ref{intro}) and the interstellar \Av{} from 0 to 40 mag \citep[cf.][]{Povich(2011)ApJS_194_14}.
Note that the SED model fitting can give very discrepant stellar and disk parameter for embedded protostars \citep{Offner(2012)ApJ_753_98,Sung(2010)AJ_140_2070}.

Table \ref{tbl-phy} lists the representative values of eight physical parameters from the SED fitting: the bolometric luminosity ($L_{bol}$), the disk accretion luminosity ($L_{acc}$), the central stellar mass ($M_{*}$), the disk mass ($M_{disk}$), the mass of envelope and ambient medium ($M_{env+amb}$), the stellar age ($t_*$), the envelope accretion rate ($\dot{M}_{acc,env}$), the disk accretion rate ($\dot{M}_{acc,disk}$), and the evolutionary stage.
$L_{acc}$ is calculated from eq. (12) of \cite{Calvet(1998)ApJ_509_802} as included in the SED fitting model \citep{Robitaille(2006)ApJS_167_256}.
The representative values are calculated referring the method used by \cite{Povich(2009)ApJ_696_1278,Povich(2011)ApJS_194_14}.
The values are the probability-weighted mean of the fitting result set $i$, whose $\chi^2_i$ satisfies $\chi^2_i-\chi^2_{min}\le2N_{data}$; $N_{data}$ is the number of data points for the SED fitting.
The probability relative to the minimum $\chi^2$ is assumed to be 
\begin{eqnarray}
P(\chi^2_i)=exp\left[-\frac{(\chi^2_i-\chi^2_{min})}{2}\right]
\end{eqnarray}
and normalized to one.
The probability-weighted mean is calculated from
\begin{eqnarray}
<X>=\sum P_iX_i
\end{eqnarray}
We calculated the probability-weighted mean in log scales, because the grid values of the model parameters spread over many orders of magnitude \citep{Robitaille(2006)ApJS_167_256}.
When the fitting model returns no disk (${M}_{disk}=0$) or no envelope (${M}_{env}=0$), we set the disk accretion rate and the envelope accretion rate as its minimum grid value, respectively (i.e. $\dot{M}_{acc,disk}=10^{-10}$ $M_{\odot}$ yr$^{-1}$ or $\dot{M}_{acc,env}=10^{-16}$ $M_{\odot}$ yr$^{-1}$) for the two parameters not to diverge in log scales.
In a similar way, we set $L_{acc}$ as its minimum value among the model set $i$, when the fitting model returns no disk.

We here note on the possible error of derived \Maccdis{}.
In the SED fitting model, \Maccdis{} is calculated from an analytic equation \citep{Whitney(2003)ApJ_591_1049,Robitaille(2006)ApJS_167_256}.
Thus, \Maccdis{} can be different from the true accretion rate.
The true accretion rate can be measured from the veiling excess as done by \cite{Hartigan(1995)ApJ_452_736}, when the central driving object is visible in optical.
In addition, the SED model fitting may give a very discrepant \Maccdis{}.
Recently, \cite{Offner(2012)ApJ_753_98} showed that \Maccdis{} from the SED fitting model can be different from the true value by a factor of $10^4$ at maximum.

\subsubsection{IFO-2 \label{dep-2}}
\Mout{} estimated from IFO-2 (HH1013 NE2) is $\sim1.8\times10^{-6}$ \Msyr{}, which is about 86 times greater than \Mout{} estimated from HH1013 NE1 using the \Ha{} intensity, $2.1\times10^{-8}$ \Msyr{} (Table \ref{tbl-out}).
\Maccdis{} of the jet-driving object is obtained to be $\sim10^{-8.93}$ \Msyr{} (Table \ref{tbl-phy}).
We then get the ratio \Mout/\Maccdis{} $\sim1.5\times10^3$ and $\sim18$ from the \Mout{} estimation using [\astFe{II}] data and \Ha{} data, respectively.
These ratios are at least $\sim150$ and $\sim2$ times higher than the ones known for YSOs \citep[$\sim10^{-2}-10^{+1}$,][]{Ellerbroek(2013)A&A_551_A5}, respectively.
\Maccenv{} is $\sim10^{-9.91}$ \Msyr.
This is smaller than \Maccdis+\Mout, hence the YSO is in the phase of losing disk mass.
$L_{bol}$ is about three orders of magnitude greater than $L_{acc}$.
$M_*$ is $\sim10^{0.4}$ \Ms{}, which corresponds to the intermediate-mass star, and $M_*+M_{disk}+M_{env+amb}\simeq2.5$ \Ms{} suggests that the source  will be an intermediate-mass star.
$t_*$ is $\sim10^{6.77}$ yr.
This age corresponds to a typical age of low- and intermediate-mass YSOs in the late evolutionary stage \citep{Andre(1994)ApJ_420_837,Bachiller(1996)ARA&A_34_111}, and it is consistent with the estimated evolutionary stage, Stage II (Table \ref{tbl-phy}), which corresponds to a T Tauri star.

\subsubsection{IFO-4 \label{dep-4}}
\Mout{} estimated from IFO-4 (HH900) is $\sim6.7\times10^{-7}$ \Msyr{}, which is about 1.2 times greater than \Mout{} estimated using the \Ha{} intensity, $5.7\times10^{-7}$ \Msyr{} (Table \ref{tbl-out}).
\Maccdis{} of the jet-driving object is obtained to be $\sim10^{-6.11}$ \Msyr{} (Table \ref{tbl-phy}).
We then get the ratio \Mout/\Maccdis{} $\sim0.86$ and $\sim0.73$ from the \Mout{} estimation using [\astFe{II}] data and \Ha{} data, respectively.
These ratios are within the range known for YSOs \citep[$\sim10^{-2}-10^{+1}$,][]{Ellerbroek(2013)A&A_551_A5}.
\Maccenv{} is $\sim10^{-4.25}$ \Msyr.
This is greater than \Maccdis+\Mout, hence the YSO is in the phase of growing disk mass.
$L_{bol}$ is about one order of magnitude greater than $L_{acc}$.
$M_*$ is $\sim10^{-0.01}$ \Ms{}, which corresponds to the low-mass star, and $M_*+M_{disk}+M_{env+amb}\simeq1.3$ \Ms{} suggests that the source will be a low-mass star.
$t_*$ is $\sim10^{3.67}$ yr.
This age corresponds to a typical age of low- and intermediate-mass YSOs in the early evolutionary stage \citep{Andre(1994)ApJ_420_837,Bachiller(1996)ARA&A_34_111}, and it is consistent with the estimated evolutionary stage, Stage 0/I (Table \ref{tbl-phy}).

\subsubsection{IFO-7 \label{dep-7}}
\Mout{} estimated from IFO-7 is $\sim9.3\times10^{-7}$ \Msyr{}, which is about 21 times greater than \Mout{} estimated from HH1014 using the \Ha{} intensity, $4.4\times10^{-8}$ \Msyr{} (Table \ref{tbl-out}).
\Maccdis{} of the jet-driving object is obtained to be $\sim10^{-6.86}$ \Msyr{} (Table \ref{tbl-phy}).
We then get the ratio \Mout/\Maccdis{} $\sim6.7$ and $\sim0.32$ from the \Mout{} estimation using [\astFe{II}] data and \Ha{} data, respectively.
These ratios are within the range known for YSOs \citep[$\sim10^{-2}-10^{+1}$,][]{Ellerbroek(2013)A&A_551_A5}, respectively.
\Maccenv{} is $\sim10^{-3.75}$ \Msyr.
This is greater than \Maccdis+\Mout, hence the YSO is in the phase of growing disk mass.
$L_{bol}$ is about 1.5 orders of magnitude greater than $L_{acc}$.
$M_*$ is $\sim10^{0.01}$ \Ms{}, which corresponds to the low-mass star, and $M_*+M_{disk}+M_{env+amb}\simeq2.8$ \Ms{} suggests that the source will be a low- or intermediate-mass star.
$t_*$ is $\sim10^{4.33}$ yr.
This age corresponds to a typical age of low- and intermediate-mass YSOs in the early evolutionary stage \citep{Andre(1994)ApJ_420_837,Bachiller(1996)ARA&A_34_111}, and it is consistent with the estimated evolutionary stage, Stage 0/I (Table \ref{tbl-phy}).

\section{Discussion \label{dis}}
\subsection{Morphology of [\astFe{II}] Jet and Outflow Features}

The morphology of the [\astFe{II}] features are knotty or longish (cf.~Fig.~\ref{fig-jet} and Table \ref{tbl-out}).
The knotty morphology may be caused by the low spatial resolution of IRIS2 images ($\sim1.5''$).
The corresponding \Ha{} features, observed with \hstacs{} of a superior imaging resolution \citep[$\sim0.1''$,][]{Smith(2010)MNRAS_405_1153}, show filamentary substructures not seen in [\astFe{II}] features (cf.~Fig.~\ref{fig-jet}).
The longish morphology probably traces the larger-scale shape of the outflow features that have finer substructures.
Most IFOs are associated with bow shocks.
A bow shock feature is clearly seen for IFO-1 and -2 in the \hstacs{} images, and could account for the appearance of IFO-3, -5, and -6b.
Especially, IFO-3 and -7 may result from a chain of bow shocks as in HH 111 \citep{Reipurth(1997)AJ_114_757}.

Among all the [\astFe{II}] features, only IFO-7 shows a clear bipolar feature.
The jet-like [\astFe{II}] feature that traces back to its driving stellar object \citep[e.g.~HH 666,][]{Smith(2004)AJ_127_2793} was observed for three IFOs (IFO-4, IFO-5, and IFO-7a) out of eight IFOs.
This shows that the [\astFe{II}] imaging is of practical use for the jet-driving source identification.

\subsection{Detection Rate of [\astFe{II}] Features \label{dis-det}}
The field we observed includes many shock-related features previously identified, such as HH objects and molecular hydrogen emission line objects \citep[MHOs,][]{Davis(2010)A&A_511_24}.
Also, numerous YSOs were identified along with their evolutionary stages.
In this section, we estimate the detection rate of IFOs in relation with this information. 

23 HH objects and 16 HH candidates were identified over several selected regions of the Carina Nebula, using ground and space telescopes \citep{Smith(2004)AJ_127_2793,Smith(2010)MNRAS_405_1153,Reiter(2013)MNRAS_433_2226}.
Among them, 10 HH objects and 5 HH candidates, including HHc-16 identified by ourselves, fall within our observation field (cf.~Fig.~\ref{fig-obs}).
We detected five IFOs that have the \Ha{} counterparts; four of them are HH objects, the remaining one is the HH candidate HHc-16 we identified.
Therefore, the [\astFe{II}] detection rate is $40$\% and $20$\% with respect to HH objects and HH candidates, respectively.
Note that this detection rate is affected not only by the shallower depth of [\astFe{II}] images but also by the poorer imaging resolution of [\astFe{II}] images.
In many cases, \Ha{} features are too close to bright point sources in the [\astFe{II}] images.

MHOs are shock-excited \Htwo{} emission line features, caused by outflows from YSOs \citep{Davis(2010)A&A_511_24}.
Six MHOs were identified in the Carina Nebula \citep{Preibisch(2011)A&A_530_A34}, and five of them fall within our observation field (Fig.~\ref{fig-obs}).
However, none of them shows [\astFe{II}] counterparts.
This non-detection is remarkable considering the $40$\% detection rate for HH objects.
This tendency was recognized from previous studies \citep[][and references therein]{Reipurth(2001)ARA&A_39_403}, and the reason seems that \Ha{} and [\astFe{II}] emissions are both bright in dissociative J-shocks, while \Htwo{} emission is bright in non-dissociative C-shocks \citep{Nisini(2002)A&A_393_1035,Reipurth(2000)AJ_120_1449,Smith(1994)A&A_289_256}.

YSOs in the Carina Nebula were identified from the SED fitting, using the photometry results from 2MASS, \spitzer{} IRAC, and \spitzer{} MIPS data \citep{Povich(2011)ApJS_194_14}.
They classified the stage of YSOs into four categories---0/I, II, III, and A (ambiguous)---according to the derived stellar mass, envelope accretion rate, and disk mass.
Table \ref{tbl-stg} lists the detection rate of IFOs with respect to the YSO stages.
Only three IFOs are determined to have definite jet-driving YSOs (cf.~section \ref{ana-res-det} and Table \ref{tbl-phy}).
Hence, these three are put in their corresponding stage row of Table \ref{tbl-stg}, while the other five IFOs whose driving YSOs are not definitely determined are only put in the total row of Table \ref{tbl-stg}.

Table \ref{tbl-stg} shows that the detection rate for the YSO evolutionary stages, however, the number of IFOs is too small to mention the statistics.
We here note that the detection rates of IFOs for class 0 and I protostars were comparable ($\sim70-80$\%) in the study of \cite{CarattioGaratti(2006)A&A_449_1077}, although their sample number is small and biased.
The total detection rate is found to be 1.4 \% (Table \ref{tbl-stg}).
This rate should be treated as a lower limit, since there can be more IFOs simply undetected due to the limit in imaging depth and imaging resolution.
For instance, the \hstacs{} data shows a higher detection rate of HH objects against the YSOs, 21/444 = 4.7 \%  \citep{Smith(2010)MNRAS_405_1153,Povich(2011)ApJS_194_14,Reiter(2013)MNRAS_433_2226}.
As far as we know, our study is the first, unbiased, wide-field [\astFe{II}] 1.64 \um{} imaging observation of jets and outflows from YSOs in the massive star forming region, where YSO's evolutionary stages are determined.
Therefore, the total detection rate of 1.4 \% in the Carina Nebula would be a reference for the future [\astFe{II}] observations of other massive star forming regions, which invokes important questions such as what factor makes the difference in the [\astFe{II}] detection rate.

\subsection{Four IFOs and Their Corresponding HH Objects}
\begin{figure}
\figurenum{2}
\label{fig-jet}
\figurenum{2a}
\center{
\includegraphics[scale=0.8]{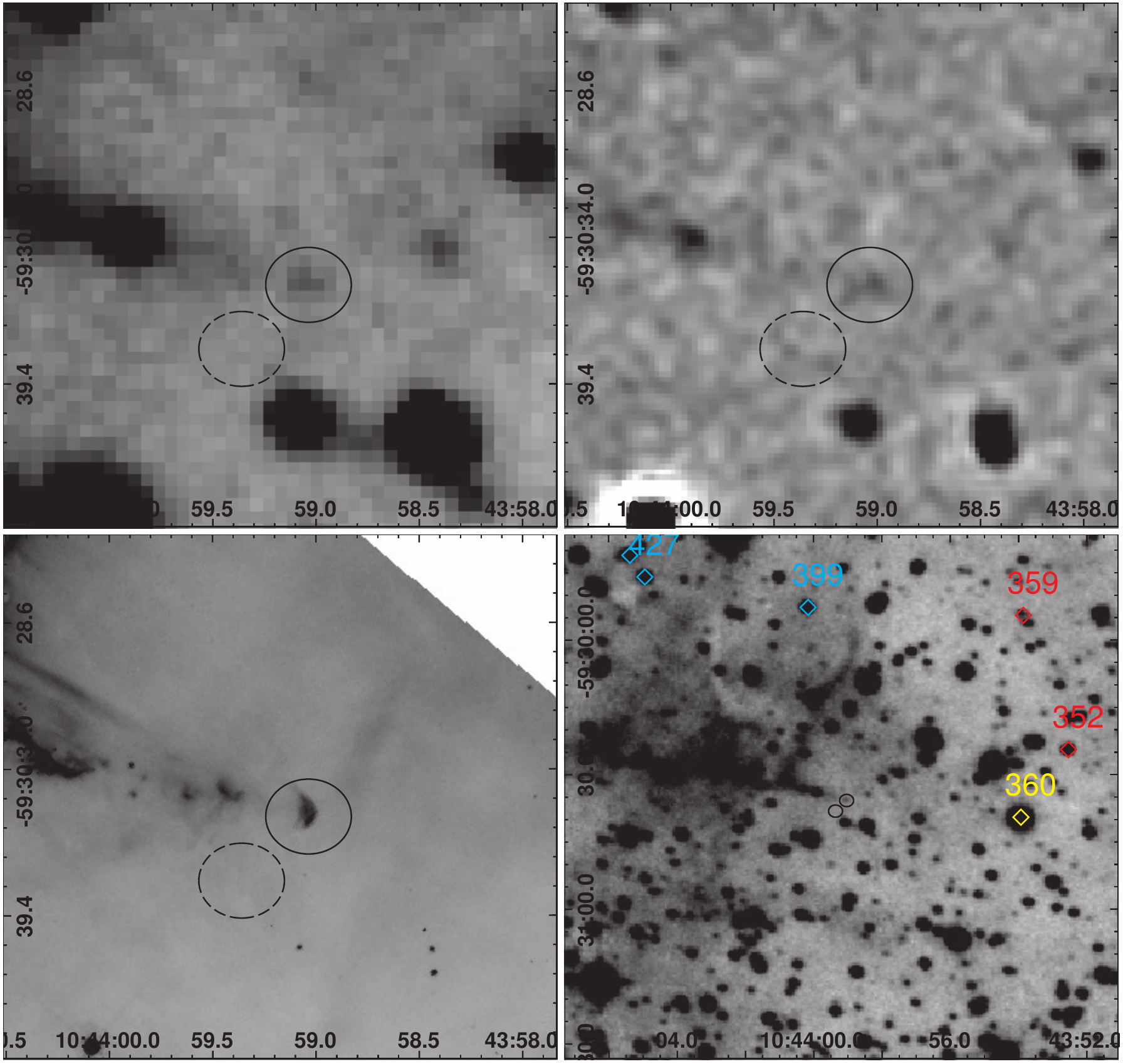}
}
\caption{The images of IFO-1 which corresponds to HH 902. The \emph{top-left} and \emph{top-right} images are the [\astFe{II}] and continuum-subtracted [\astFe{II}] images obtained from \aatiris, respectively. The \emph{bottom-left} image is the F658N (\Ha{}+[\astN{II}]) filter image obtained from \hstacs{} \citep[cf.][]{Smith(2010)MNRAS_405_1153}. The \emph{bottom-right} image is the zoomed-out [\astFe{II}] image. The \emph{solid-ellipse and dashed-ellipse} are the source and background area used for the [\astFe{II}] flux measurement, respectively (cf.~Table \ref{tbl-flux}). The diamonds indicate the positions of YSOs identified by \cite{Povich(2011)ApJS_194_14} in the stage of 0/I (\emph{red}), II (\emph{yellow}), III (\emph{green}), or ambiguous (\emph{cyan}), and the corresponding number is the catalog number. The coordinates are RA and Dec in J2000. \label{fig-jet-a}}
\end{figure}

\begin{figure}
\figurenum{2c}
\center{
\includegraphics[scale=0.8]{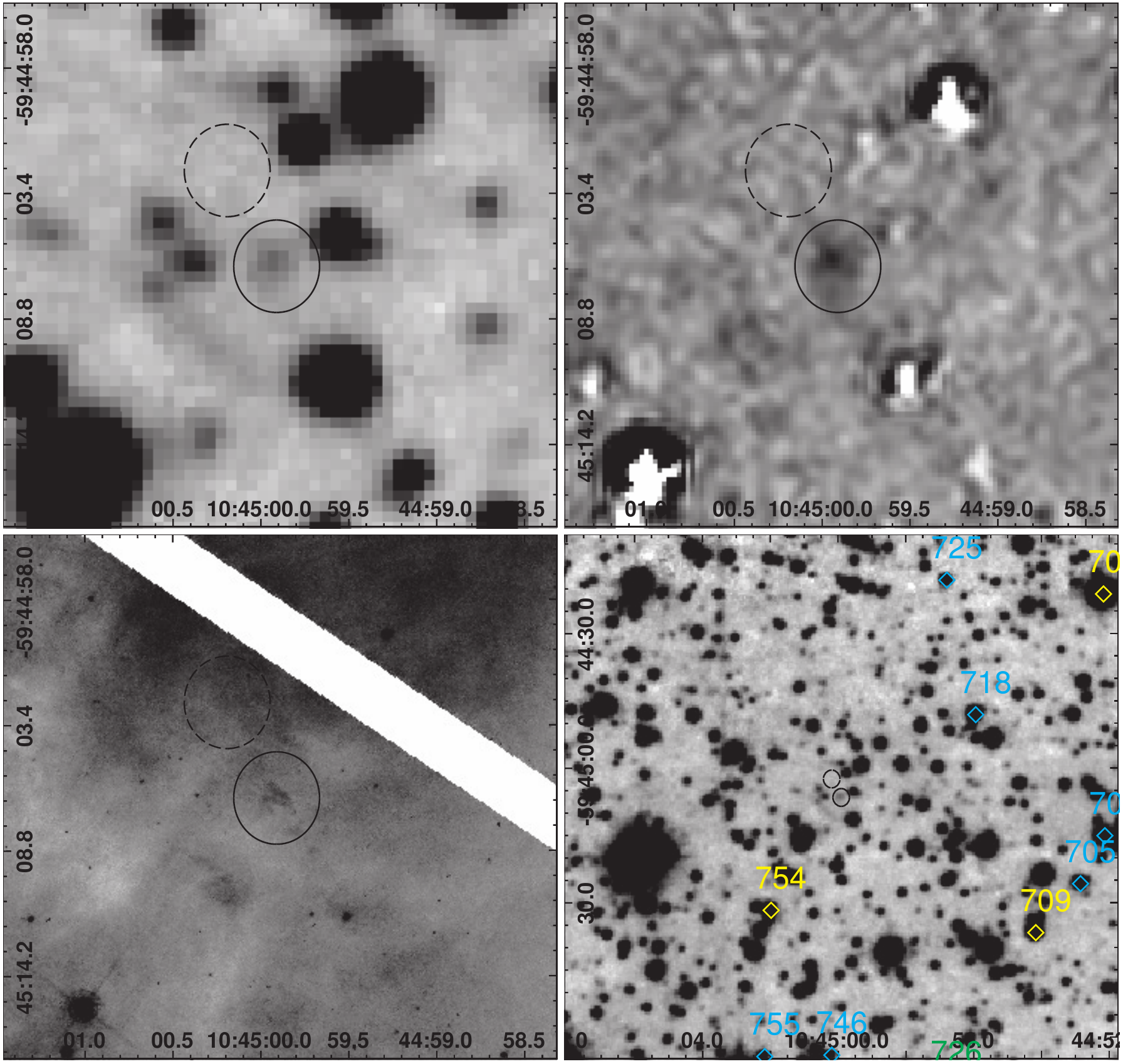}
}
\caption{The images of IFO-3 which corresponds to the newly-discovered HHc-16. The rest is the same as for Figure \ref{fig-jet-a}. \label{fig-jet-c}}
\end{figure}

\begin{figure}
\figurenum{2d}
\center{
\includegraphics[scale=0.8]{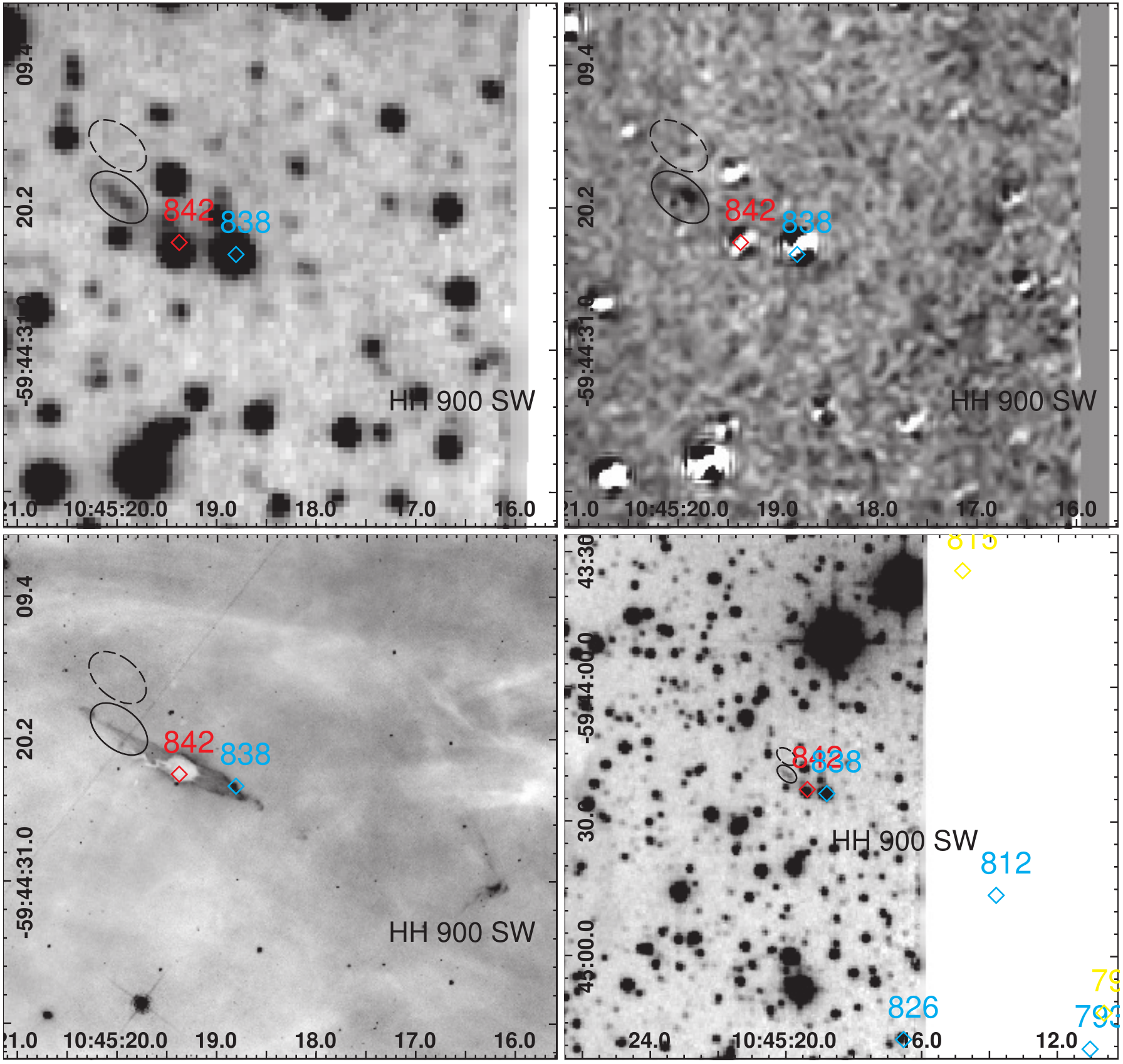}
}
\caption{The images of IFO-4 which corresponds to HH 900. The \aatiris{} image was obtained while observing the T4 region (Fig.~\ref{fig-obs}). The rest is the same as for Figure \ref{fig-jet-a}. \label{fig-jet-d}}
\end{figure}

\begin{figure}
\figurenum{2e}
\center{
\includegraphics[scale=0.8]{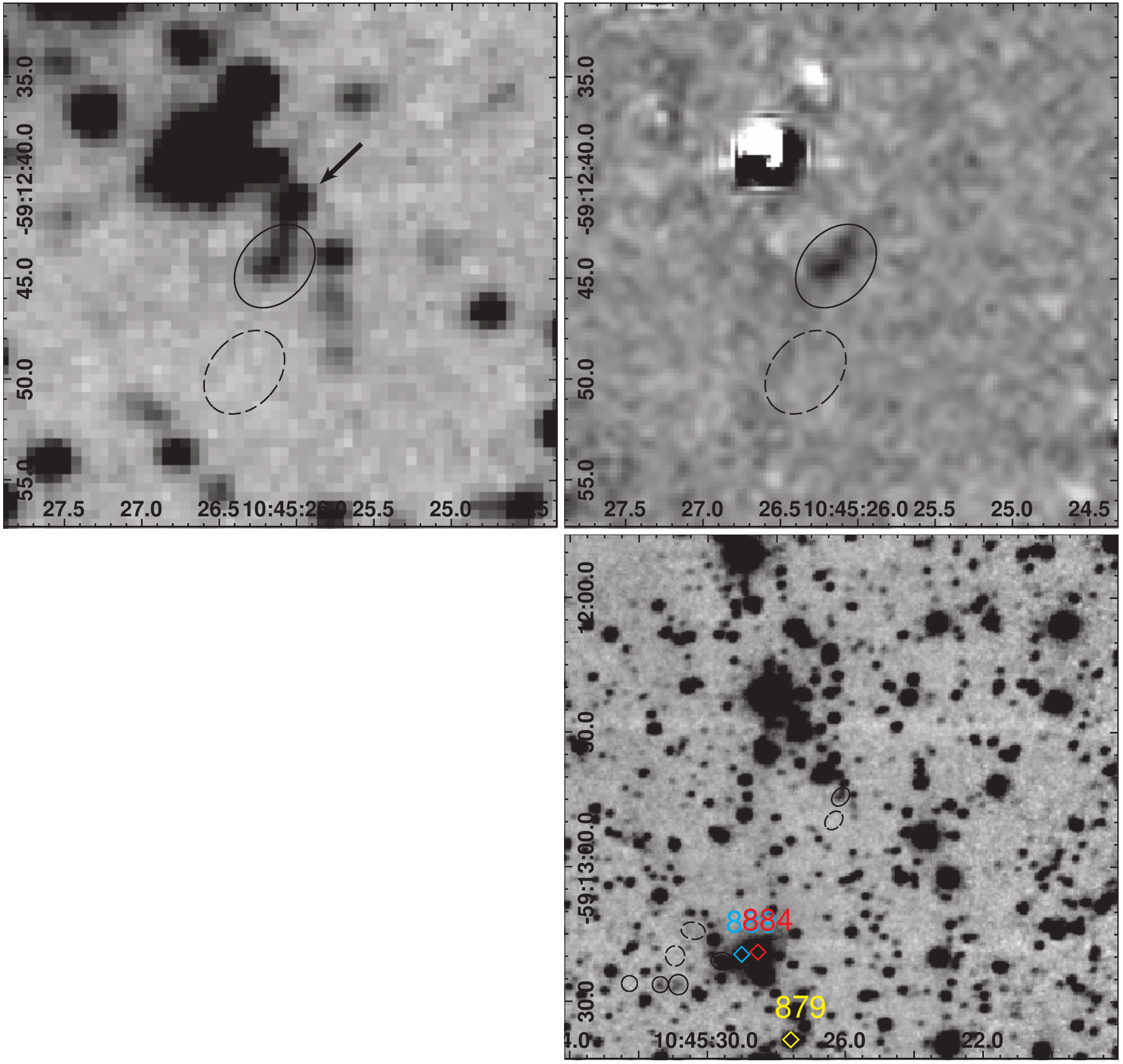}
}
\caption{The images of IFO-5. The rest is the same as for Figure \ref{fig-jet-a}. The black arrow in the upper-left panel indicates a jet-driving candidate (see text). \label{fig-jet-e}}
\end{figure}

\begin{figure}
\figurenum{2f}
\center{
\includegraphics[scale=0.8]{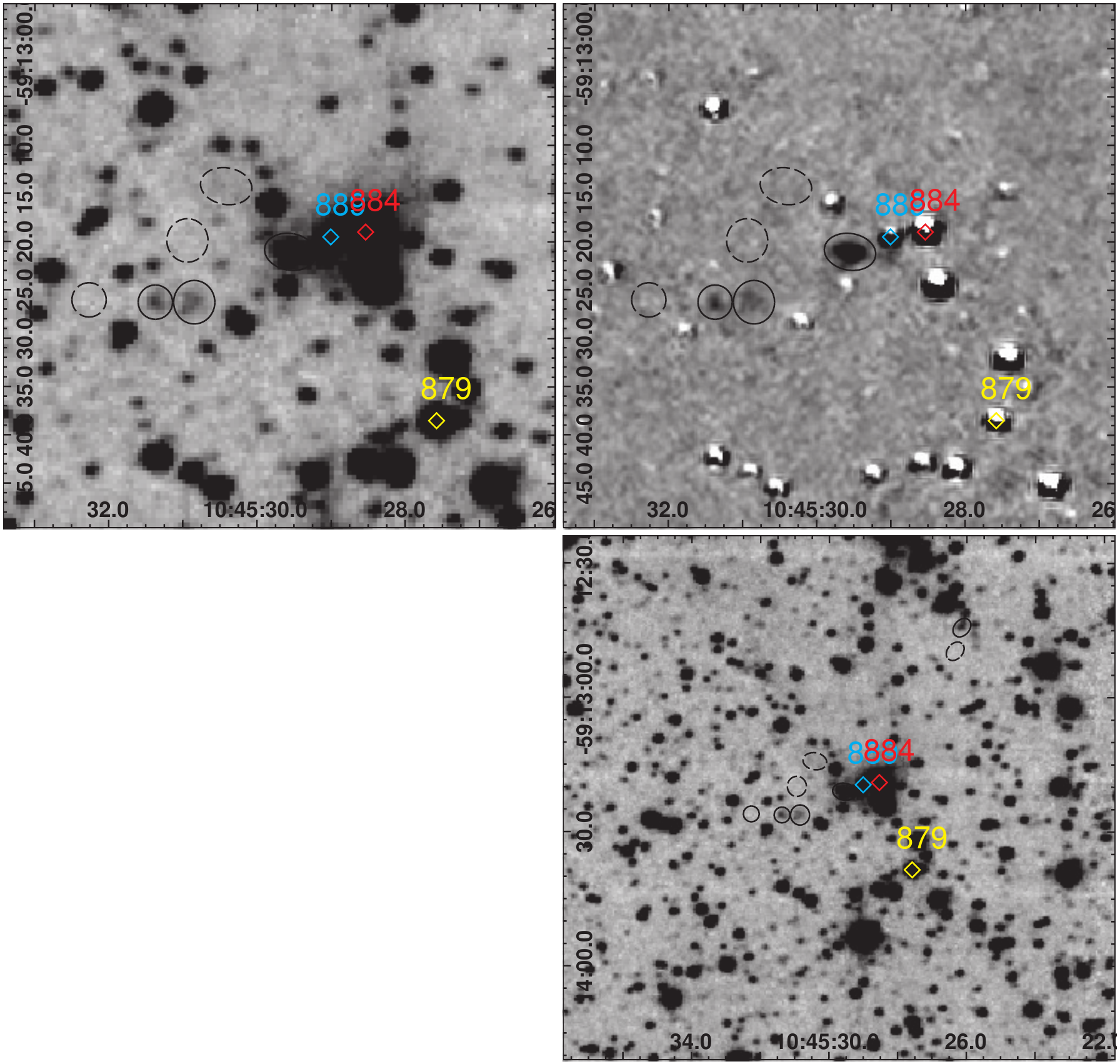}
}
\caption{The images of IFO-6a, IFO-6b, and IFO-6c (from right to left). The rest is the same as for Figure \ref{fig-jet-a}. \label{fig-jet-f}}
\end{figure}

\begin{figure}
\figurenum{2g}
\center{
\includegraphics[scale=0.8]{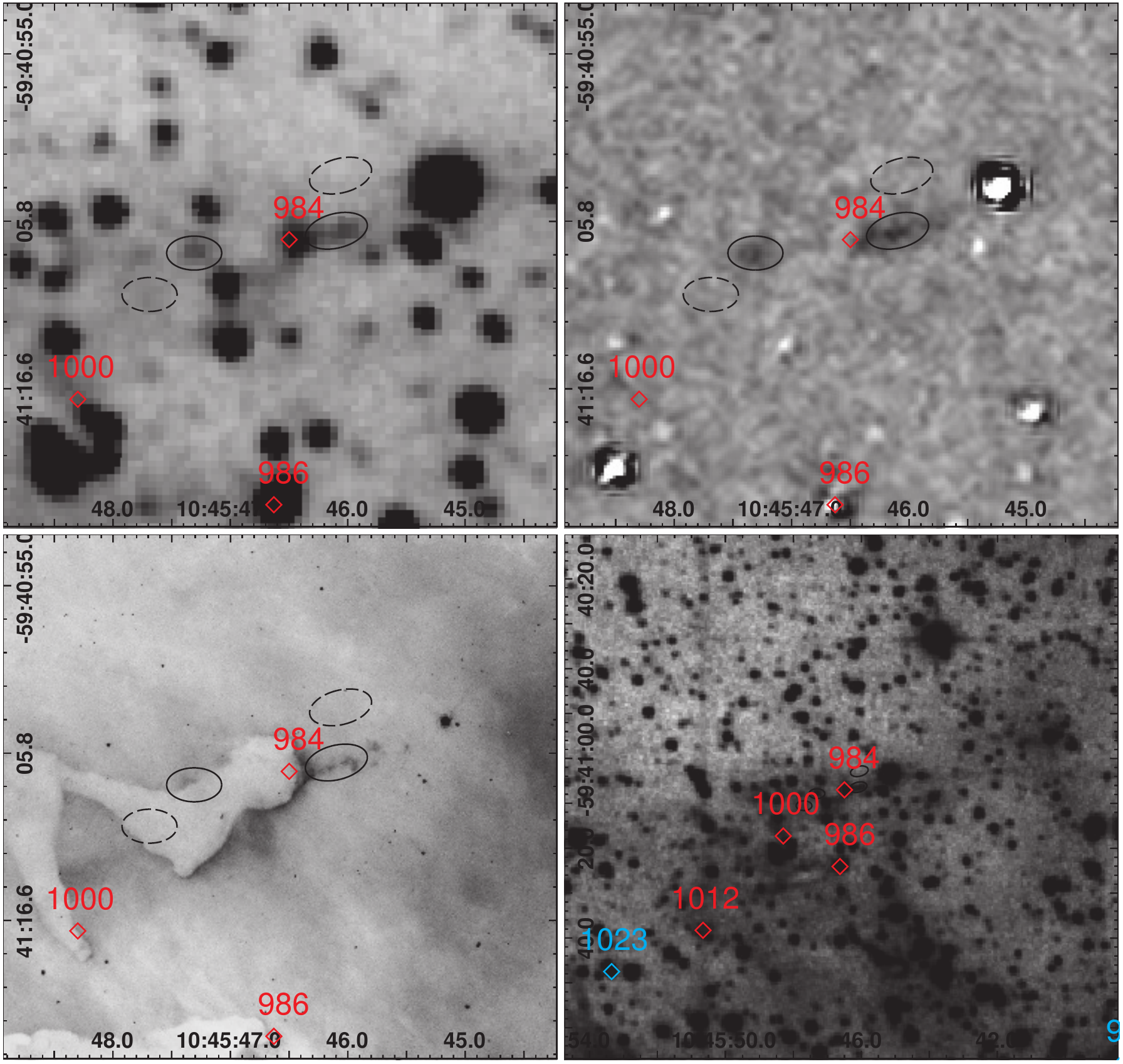}
}
\caption{The images of IFO-7a and IFO-7b (from right to left) which corresponds to HH 1014. The rest is the same as for Figure \ref{fig-jet-a}. \label{fig-jet-g}}
\end{figure}

\begin{figure}
\figurenum{2h}
\center{
\includegraphics[scale=0.8]{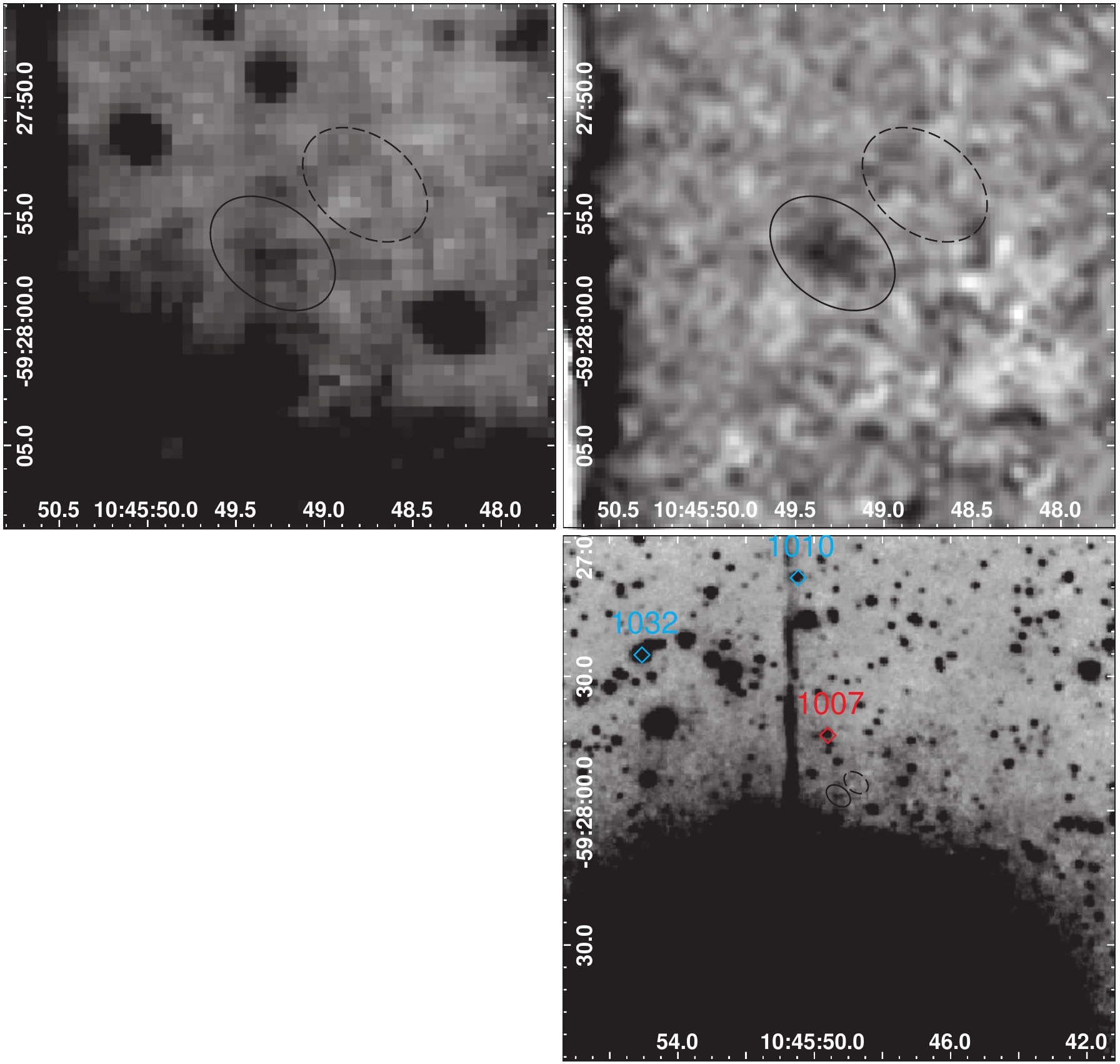}
}
\caption{The images of IFO-8. The rest is the same as for Figure \ref{fig-jet-a}. The bright star at the southeastern corner is BO Car (cf.~Fig.~\ref{fig-obs}). \label{fig-jet-h}}
\end{figure}

\begin{figure}
\figurenum{4}
\center{
\includegraphics[scale=0.8]{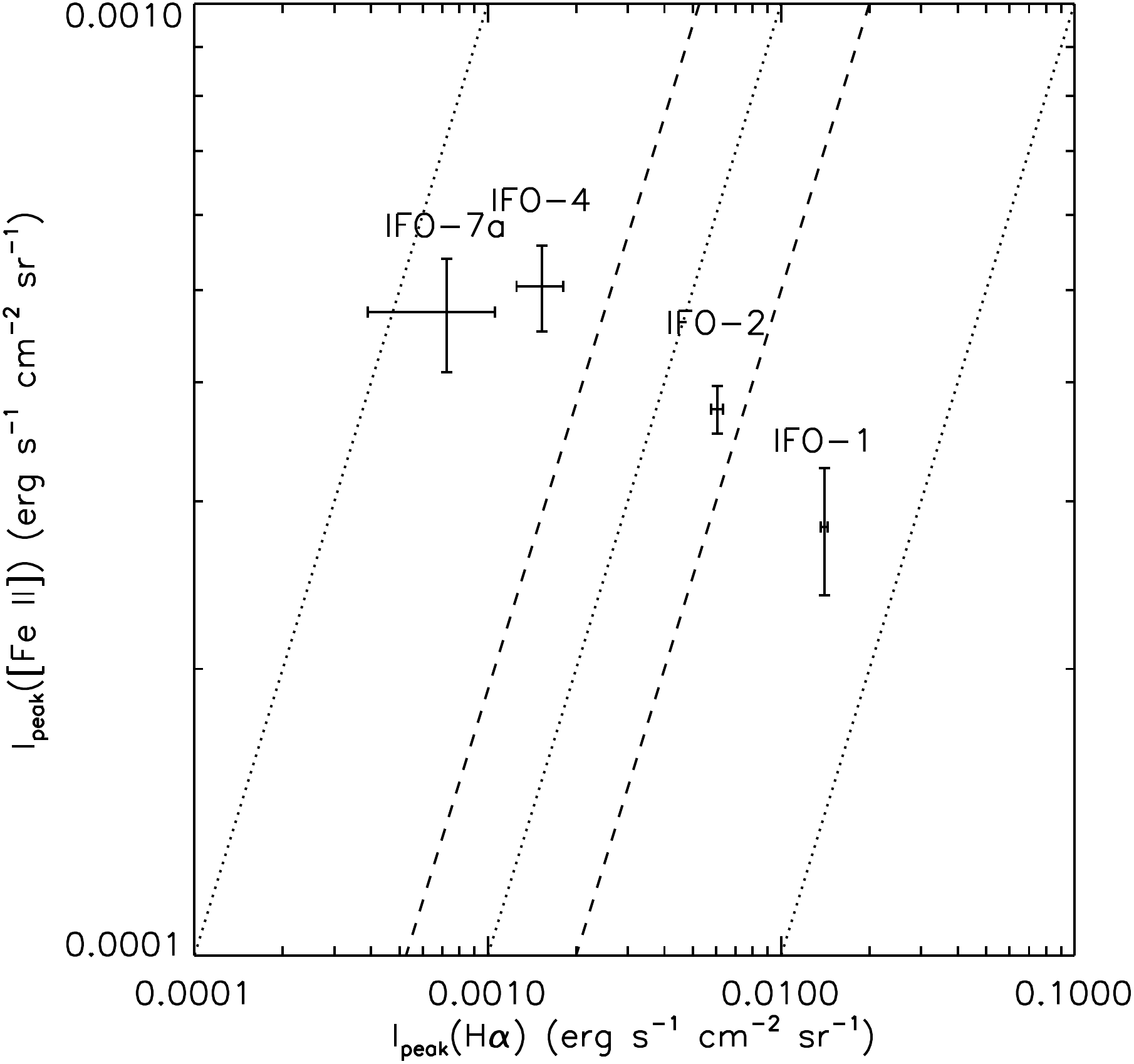}
}
\caption{Plot of IFO peak intensities with respect to their corresponding \Ha{} intensities. The \Ha{} intensity is measured at the [\astFe{II}] peak position from the \hstacs{} data. Both intensities are extinction-corrected. The \emph{dotted-lines} indicates the ratio of [\astFe{II}]/\Ha=1, 0.1, 0.01 from left to right. The \emph{dashed-lines} indicates the ratio of [\astFe{II}]/\Ha=0.19 and 0.05 from left to right. IFO-2 and -4 are the averages of the values from the two overlapped regions, respectively (cf.~Fig.~\ref{fig-obs} and text). \label{fig-feii-ha}}
\end{figure}

As mentioned in section \ref{dis-det}, only four IFOs have the corresponding HH objects; IFO-1, -2, -4, and -7a correspond to HH 902, 1013 NE2, 900, and 1014, respectively (cf.~Table \ref{tbl-flux} and Fig.~\ref{fig-jet}).
In order to obtain some hints of IFO's physical properties, we compared the peak intensity of IFO in [\astFe{II}] and \Ha. 
We picked the position of [\astFe{II}] peak, and at this position the average \Ha{} intensities were measured from the \hstacs{} data \citep{Smith(2010)MNRAS_405_1153}.
The measured intensities were then extinction-corrected as described in section \ref{ana-res-flux}. 
$A_{\textrm{\Ha}}$ is about 2.54.
Figure \ref{fig-feii-ha} is the resultant plot of intensity comparison.
For IFO-2 and -4, we have averaged the values from the two overlapped regions (cf.~section \ref{ana-res-det-2} and \ref{ana-res-det-4}).

The ratio [\astFe{II}]/\Ha{} ranges from $\sim0.01$ to $\sim1.0$ (Fig.~\ref{fig-feii-ha}).
The [\astFe{II}] and \Ha{} intensities shows a high anti-correlation (Pearson correlation coefficient of $-0.93$), although the number of points are small.
In a dissociative J-shock model \citep{Allen(2008)ApJS_178_20}, [\astFe{II}]/\Ha{} ranges $0.05-0.19$ for shock velocities of $100-300$ \kms, preshock densities of $1-10^3$ \ncm, and solar abundance.
Only IFO-2 falls within this range, and the rest has a lower or higher ratio.
We here note that the ``knotty'' IFOs have a lower [\astFe{II}]/\Ha{} than ``longish'' IFOs (Table \ref{tbl-out}).
The ``knotty'' IFOs (IFO-1 and -2) have a bow shock feature in the \Ha{} image, while the ``longish'' IFOs (IFO-4 and -7a) have a jet-body-like feature (Fig.~\ref{fig-jet}).
This morphological difference may be related with the difference in [\astFe{II}]/\Ha{} and in the dominant excitation mechanism for [\astFe{II}] and \Ha{} features (e.g.~shock or photo-ionization).
Line ratio studies through spectroscopic observations would provide more accurate information on these properties.
Emission lines that are close in wavelength are desirable, because the line ratio is then less dependent on extinction.

\begin{figure}
\figurenum{2b}
\center{
\includegraphics[scale=0.8]{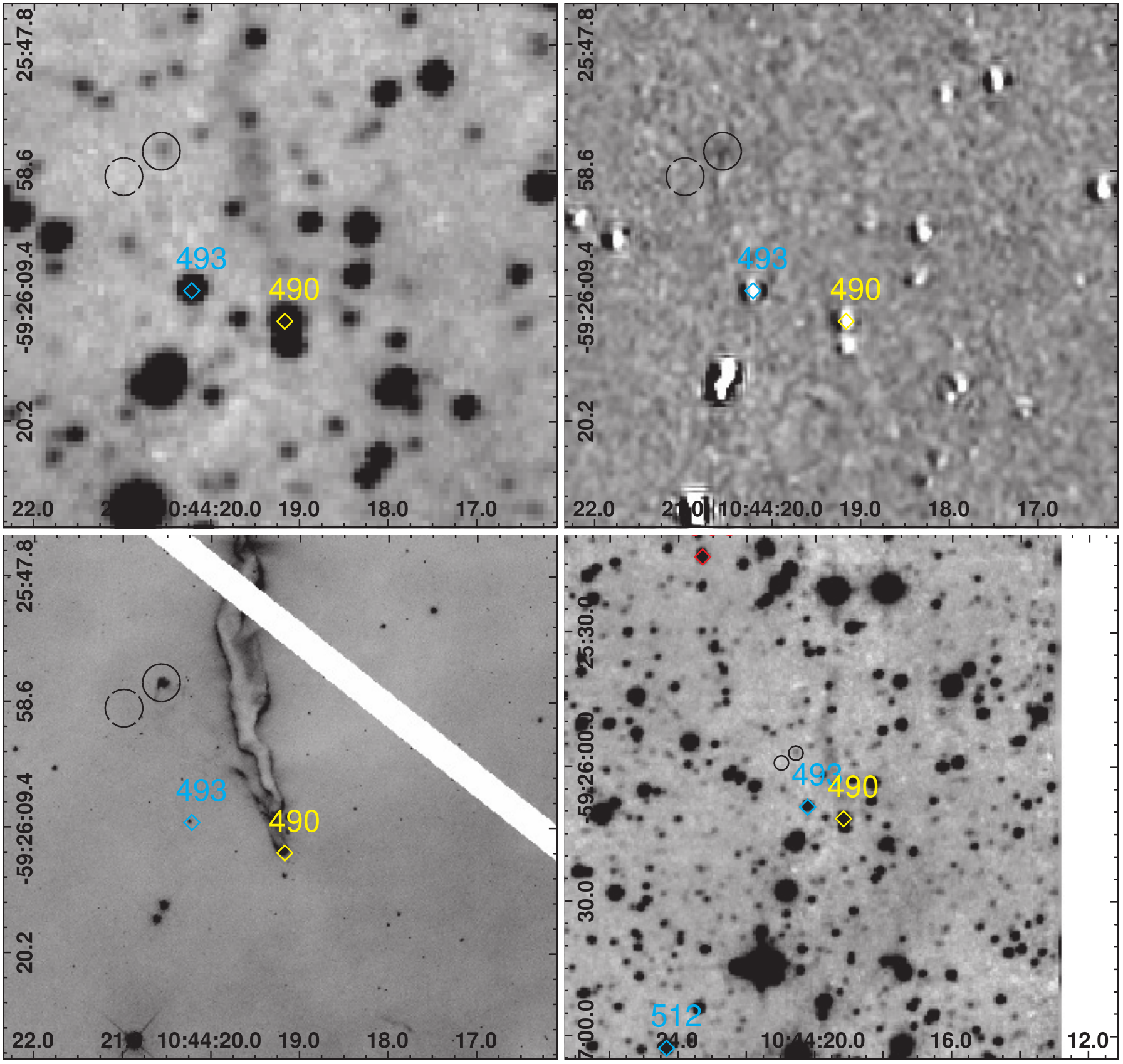}
}
\caption{The images of IFO-2 which corresponds to HH 1013 NE2. The \aatiris{} image was obtained while observing the S2 region (Fig.~\ref{fig-obs}). The rest is the same as for Figure \ref{fig-jet-a}. \label{fig-jet-b}}
\end{figure}

\Mout{} estimated from [\astFe{II}] and \Ha{} data is different within one or two orders of magnitude (Table \ref{tbl-out}).
\Mout{} from [\astFe{II}] is greater than \Mout{} from \Ha{} data, although it is not that much greater for IFO-4.
\cite{Smith(2010)MNRAS_405_1153} did not perform the extinction correction in estimating \Mout{} from \Ha{} data.
If $A_{\textrm{\Ha}}=2.54$ mentioned above is applied, \Mout{} from \Ha{} data is increased by about a factor of three, since \Mout{} is proportional to the square-root of \Ha{} intensity in the estimation of \cite{Smith(2010)MNRAS_405_1153}. 
The neutral gas excluded in the \Ha{} estimation is another reason for the different \Mout{} \citep{Reiter(2013)MNRAS_433_2226}. 
For IFO-1 and -2, the different dynamical time, $\tau$, used in section \ref{out-long} could account for the difference, because the [\astFe{II}] feature used for the \Mout{} estimation is farther from the jet-driving object than the \Ha{} feature used for the \Mout{} estimation (Fig.~\ref{fig-jet-a}, \ref{fig-jet-b}, and \citealt{Smith(2010)MNRAS_405_1153}).
In addition, an episodic, non-steady outflowing may contribute to the different \Mout, since \Mout{} of a knotty IFO is estimated at the instance of shock excitation rather than averaging over a time duration.
The outflow velocity is not likely the factor that makes the difference in the outflow mass loss rate estimation, because the same velocity was used for the two estimations, i.e.~200 \kms.
The several assumptions used for both estimations hamper clarifying what dominantly causes the different \Mout{} (cf.~section \ref{out-cav} and \citealt{Smith(2010)MNRAS_405_1153}).
Spectroscopic observations on the jet-driving object would give more accurate, current \Mout.

\subsection{Three In-depth Inspected IFOs: IFO-2, -4, and -7}

We have more deeply investigated three IFOs whose jet-driving objects are identified in terms of the outflow mass loss rate and the physical characteristics of the jet-driving object (cf.~section \ref{ana-res-dep}).
In this section, the general features seen in the relevant physical parameters and their mutual relations are discussed.

\Mout/\Maccdis{} of YSOs from previous observational studies shows that it ranges $\sim10^{-2}-10^{+1}$, according to the compilation by \cite{Ellerbroek(2013)A&A_551_A5}.
The compilation is from updated estimations using more reliable diagnostics (private communication with L. Ellerbroek in 2012).
We note that this range covers higher ratios than the frequently referred ratio of $\sim0.1$ estimated from classical T Tauri stars \citep{Ray(2007)2007prpl.conf__231}.
Specifically speaking, the majority of the objects that shows \Mout/\Maccdis{} $\sim1.0$ is Herbig Ae/Be stars; however, classical T Tauri stars and Class 0/I YSOs do show \Mout/\Maccdis{} $>1.0$ in some cases.
HN Tau (T Tauri star) and T Tau (Class 0/I) are the examples shown in \cite{Ellerbroek(2013)A&A_551_A5}.
\cite{Ellerbroek(2013)A&A_551_A5} used the values from literatures as follows: \Mout(HN Tau) of $10^{-8.10}$ \Msyr\ \citep{Hartigan(1995)ApJ_452_736}, \Maccdis(HN Tau) of $10^{-8.89}$ \Msyr\ \citep{Muzerolle(1998)AJ_116_2965}, \Mout(T Tau) of $10^{-6.40}$ \Msyr\ \citep{Podio(2012)A&A_545_A44,Herbst(1997)AJ_114_744}, and \Maccdis(T Tau) of $10^{-7.05}$ \Msyr\ \citep{Podio(2012)A&A_545_A44,Calvet(2004)AJ_128_1294}.

\Mout/\Maccdis{} from our estimation falls within the range \cite[$\sim10^{-2}-10^{+1}$,][]{Ellerbroek(2013)A&A_551_A5} for IFO-4 and -7, but exceeds it for IFO-2.
IFO-4 and -7 have \Mout/\Maccdis{} of $\sim0.32-6.7$ (section \ref{dep-4} and \ref{dep-7}).
Considering this \Mout/\Maccdis{} with their YSO stage of 0/I (Table \ref{tbl-phy}), the jet-driving objects are likely to be Class 0/I YSOs.
IFO-2 has \Mout/\Maccdis{} of $\sim1.5\times10^3$ or $\sim18$, depending on which \Mout{} estimation is used ([\astFe{II}] or \Ha, cf.~section \ref{dep-2}).
These estimations exceed the range \cite[$\sim10^{-2}-10^{+1}$,][]{Ellerbroek(2013)A&A_551_A5} by a several factor or up to $\sim100$ factor.

This excess may be caused by either the overestimation of \Mout{} (cf.~section \ref{out-cav}) or the underestimation of \Maccdis{} from the SED fitting model (cf.~section \ref{sed-fit}).
The \Mout{} of IFO-2 is not exceptionally higher than other IFOs (Table \ref{tbl-out}), hence we think that the overestimation of \Mout{} is less plausible.
A severe underestimation of \Maccdis{} is possible by erroneous estimation of SED model fitting.
We here note that \Maccdis{} from the model can be different from the true value as much as a factor of $\sim10^4$ \citep{Offner(2012)ApJ_753_98}.
Another possibility is the misidentification of the jet-driving object.
The YSO \#490 is slightly misaligned with the probable jet-axis inferred from the \Ha{} bow shock shape (Fig.~\ref{fig-jet}), although \#490 can have a relative motion to IFO-2 due to radiation pressure, for example.
If IFO-2 is generated by another YSO in an earlier evolutionary stage than \#490, \Maccdis{} can be higher and hence make \Mout/\Maccdis{} lower.
The proper motion study of IFO-2 would help identifying of jet-driving object.

The other physical quantities show reasonable relations or trends. 
\Maccenv{} is greater than \Maccdis{}+\Mout{} for IFO-4 and -7, while it is reversed for IFO-2.
This means that the disks of IFO-4 and -7 are gaining their mass, while the disk of IFO-2 is losing its mass.
This trend seems to be consistent with the inferred evolutionary stage of IFO-2 (Stage II) and IFO-4 and -7 (Stage 0/I) (cf.~Table \ref{tbl-phy}).
The ratio of $L_{bol}/L_{acc}$ is the highest for IFO-2 ($\sim10^3$), while it is much smaller for IFO-4 and -7 ($\sim10^{1.0}-10^{1.5}$).
This also seems to be compatible with the inferred evolutionary stages (Table \ref{tbl-phy}).
A larger contribution of $L_{acc}$ in $L_{bol}$ is usually expected at the early evolutionary stage.
As $M_*$, $M_{disk}$, and $M_{env+amb}$ show (Table \ref{tbl-phy}), the three YSOs are likely to be low- or intermediate-mass stars.
$t_*$ shows two age scales.
One is $t_*\sim10^{3}-10^{4}$ yr, and the other is $t_*\sim10^{7}$ yr (Table \ref{tbl-phy}).
These ages are consistent with the typical ages for YSOs in the early and late evolutionary stages, respectively  \citep{Andre(1994)ApJ_420_837,Bachiller(1996)ARA&A_34_111}. 
The age interval is also consistent with the age spread of YSOs in the Carina Nebula inferred by \cite{Povich(2011)ApJS_194_14}, $\sim2-5$ Myr.

\section{Conclusion}
We performed [\astFe{II}] 1.64 \um{} imaging observations over the northern part ($\sim24'\times45'$) of the Carina Nebula with IRIS2 of Anglo-Australian Telescope.
Eleven jet and outflow features, named as IFOs, were detected at eight different regions, and four of those features correspond to HH objects previously identified from \hstacs{} observations \citep{Smith(2010)MNRAS_405_1153}.
One HH candidate is newly identified by comparing our [\astFe{II}] images to the \Ha{} image of \hstacs.

The morphology of IFOs is knotty or longish, but all of them are thought to have unresolved, possibly bow shaped, substructures.
IFOs show 40 \% and 0 \% detection rate for HH objects and MHOs, which seems to reflect the different shock origins of [\astFe{II}] and \Htwo{} features (dissociative and non-dissociative shocks).
The detection rate of IFOs against YSOs in the Carina Nebula is estimated to be $\sim1.4$ \%, which should be a lower limit since many IFOs may be missed due to the limits in imaging depth and spatial resolution. 
Four IFOs (IFO-1, 2, 4, and 7a) have their relevant HH objects.
Their peak intensities in [\astFe{II}] and \Ha{} show a high anti-correlation, with the measured intensity ratios spanning the range expected for shocks.
These properties may be related with the IFO morphologies, and subsequently excitation mechanisms (e.g.~shock or photo-ionization).

\Mout{} is estimated from [\astFe{II}] flux using two different methods, and it ranges $\sim10^{-7}-10^{-6}$ \Msyr{} (Table \ref{tbl-out}).
We identified the jet-driving objects for three IFOs (IFO-2, -4, and -7).
For these three IFOs, we compared the relations between \Mout{} and the physical parameters of jet-driving objects, which are derived from the SED model fitting.
The ratios \Mout/\Maccdis{} are consistent for IFO-4 and -7 with values quoted in the literature \citep[$\sim10^{-2}-10^{+1}$,][]{Ellerbroek(2013)A&A_551_A5}, while the estimated ratio for IFO-2 is greater by a several factor or up to $\sim100$ factor.
This excess for IFO-2 seems to stem from the underestimation of \Maccdis{}, since \Mout{} of IFO-2 is not exceptionally higher than others.
The causes of the underestimation include the discrepant \Maccdis{} from the SED fitting model and the misidentification of the jet-driving object.
The differences between \Maccdis{}+\Mout{} and \Maccenv{}, and between $L_{bol}$ and $L_{acc}$ show trends that are consistent with their evolutionary stages given from the SED model fitting.
$M_*$ are of low- or intermediate-mass stars.
$t_*$ shows two distinctive values of $\sim10^3-10^4$ yr and $\sim10^7$ yr, which are consistent with the corresponding evolutionary stage and the age spread of YSOs in the Carina Nebula \citep{Povich(2011)ApJS_194_14}.

\acknowledgments
J.-H.S. and T.-S.P. express their gratitude to the anonymous referee for numerous useful comments and to Bo Reipurth for the discussion on the identification of HH objects.
J.-H. S. is grateful to Matthew Povich and Kwang-Il Seon for the discussion about the SED fitting and to Lucas Ellerbroek for the target information.
This research has been supported by K-GMT Science Program, and made use of SAOImage DS9, developed by Smithsonian Astrophysical Observatory \citep{Joye(2003)inproc}.
B.-C. K. was supported by the National Research Foundation of Korea (NRF) grant funded by the Korea Government (MEST) (No. 2012R1A4A1028713).


\bibliographystyle{../_bibtex/bst/apj}
\bibliography{../_bibtex/paper,../_bibtex/book,../_bibtex/manual,../_bibtex/megazine}

\clearpage

\clearpage

\clearpage

\clearpage

\clearpage

\clearpage

\clearpage

\clearpage

\clearpage

\clearpage

\clearpage

\clearpage
\begin{deluxetable}{cccccccc}
\tabletypesize{\footnotesize}
\tablecaption{Summary of the \aatiris{} Imaging Observations \label{tbl-obs}}
\tablehead{
\colhead{Region} &  \colhead{Pointing Position} & \colhead{Date} & \colhead{Filter} & \colhead{Dithering} & \colhead{Cycle} & \colhead{Period} & \colhead{Total Exposure} \\
& \colhead{(RA, Dec; J2000)} & \colhead{(in 2011)} & & & & \colhead{(s)} & \colhead{(s)} }
\startdata
T1& 10:45:50.70, $-$59:49:53.8 & Mar-27 & [\astFe{II}] & 9 & 10 & 20 & 1800 \\
&&& H & 9 & 10 & 1 & 90 \\
T2& 10:44:51.17, $-$59:48:07.7 & Feb-22 & [\astFe{II}] & 9 & 10 & 20 & 1800 \\
&&& H & 9 & 10 & 1 & 90 \\
T3& 10:43:52.54, $-$59:50:08.2 & Feb-24 & [\astFe{II}] & 10 & 10 & 20 & 2000 \\
&&& H & 9 & 10 & 1 & 90 \\
T4& 10:45:50.18, $-$59:42:36.2 & Mar-26 & [\astFe{II}] & 9 & 10 & 20 & 1800 \\
&&& H & 9 & 10 & 1 & 90 \\
T5& 10:44:51.23, $-$59:40:52.7 & Feb-22 & [\astFe{II}] & 9 & 10 & 20 & 1800 \\
&&& H & 9 & 10 & 1 & 90 \\
T6& 10:43:52.53, $-$59:42:55.6 & Feb-24 & [\astFe{II}] & 9 & 10 & 20 & 1800 \\
&&& H & 9 & 10 & 1 & 90 \\
T7& 10:45:48.65, $-$59:36:21.7 & Feb-24 & [\astFe{II}] & 9 & 10 & 20 & 1800 \\
&&& H & 9 & 10 & 1 & 90 \\
T8& 10:44:51.07, $-$59:33:40.0 & Feb-23 & [\astFe{II}] & 9 & 10 & 20 & 1800 \\
&&& H & 9 & 10 & 1 & 90 \\
T9& 10:43:53.34, $-$59:35:45.0 & Feb-23 & [\astFe{II}] & 9 & 10 & 20 & 1800 \\
&&& H & 9 & 10 & 1 & 90 \\
S1& 10:45:48.53, $-$59:29:10.1 & Feb-24 & [\astFe{II}] & 9 & 20 & 10 & 1800 \\
&&& H & 9 & 10 & 1 & 90 \\
S2& 10:44:51.02, $-$59:26:24.6 & Feb-24 & [\astFe{II}] & 9 & 20 & 10 & 1800 \\
&&& H & 9 & 10 & 1 & 90 \\
S3& 10:43:52.67, $-$59:28:32.0 & Feb-23 & [\astFe{II}] & 9 & 10 & 20 & 1800 \\
&&& H & 9 & 1 & 10 & 90 \\
S4& 10:45:49.53, $-$59:20:55.4 & Mar-26 & [\astFe{II}] & 10 & 10 & 20 & 2000 \\
&&& H & 9 & 10 & 1 & 90 \\
S5& 10:44:51.08, $-$59:19:12.1 & Feb-22 & [\astFe{II}] & 10 & 10 & 20 & 2000 \\
&&& H & 9 & 10 & 1 & 90 \\
S6& 10:43:53.23, $-$59:21:15.5 & Feb-23 & [\astFe{II}] & 10 & 10 & 20 & 2000 \\
&&& H & 9 & 10 & 1 & 90 \\
S7& 10:45:47.95, $-$59:14:38.3 & Feb-24 & [\astFe{II}] & 10 & 10 & 20 & 2000 \\
&&& H & 9 & 10 & 1 & 90 \\
S8& 10:44:51.00, $-$59:11:59.3 & Feb-23 & [\astFe{II}] & 10 & 10 & 20 & 2000 \\
&&& H & 9 & 10 & 1 & 90 \\
S9& 10:43:52.56, $-$59:14:03.4 & Feb-23 & [\astFe{II}] & 10 & 10 & 20 & 2000 \\
&&& H & 9 & 1 & 10 & 90 \\

\enddata
\end{deluxetable}

\clearpage
\begin{deluxetable}{cccrr}
\tablewidth{0pt}
\tablecaption{[\astFe{II}] Flux Measurements \label{tbl-flux}}
\tablehead{
\colhead{Targets} & \colhead{RA, Dec (J2000)} & \colhead{HH} & \multicolumn{2}{c}{[Fe II] flux} \\
& & \colhead{Object\tablenotemark{\dag1}} & \colhead{Observed} & \colhead{Dereddened\tablenotemark{\dag2}} \\
& & & \multicolumn{2}{c}{($10^{-15}$ erg s$^{-1}$ cm$^{-2}$)}
}
\startdata
                         IFO-1 &   10:43:59.036, $-$59:30:35.77 &      HH902 &   8.8$\pm$1.5 &  15.1$\pm$2.6 \\
    IFO-2\tablenotemark{\dag3} &   10:44:20.566, $-$59:25:57.00 & HH1013 NE2 &  13.8$\pm$1.3 &  23.6$\pm$2.3 \\
                         IFO-3 &   10:44:59.910, $-$59:45:06.55 &    \nodata &  15.4$\pm$1.5 &  26.4$\pm$2.5 \\
    IFO-4\tablenotemark{\dag4} &   10:45:19.854, $-$59:44:19.98 &      HH900 &  37.2$\pm$1.8 &  63.8$\pm$3.2 \\
                         IFO-5 &   10:45:26.143, $-$59:12:44.38 &    \nodata &  23.6$\pm$1.1 &  40.4$\pm$2.0 \\
                        IFO-6a &   10:45:29.555, $-$59:13:21.05 &    \nodata &  93.6$\pm$4.0 & 160.0$\pm$6.8 \\
                        IFO-6b &   10:45:30.849, $-$59:13:26.27 &    \nodata &  17.1$\pm$1.0 &  29.3$\pm$1.7 \\
                        IFO-6c &   10:45:31.373, $-$59:13:26.27 &    \nodata &  12.6$\pm$0.9 &  21.6$\pm$1.5 \\
                        IFO-7a &   10:45:46.098, $-$59:41:06.43 &     HH1014 &  32.6$\pm$2.3 &  55.8$\pm$4.0 \\
                        IFO-7b &   10:45:47.309, $-$59:41:07.87 &    \nodata &   8.7$\pm$2.0 &  14.9$\pm$3.5 \\
                         IFO-8 &   10:45:49.292, $-$59:27:56.72 &    \nodata &  38.7$\pm$4.9 &  66.2$\pm$8.4 \\

\enddata
\tablenotetext{\dag1}{The corresponding HHO discovered by \cite{Smith(2010)MNRAS_405_1153}.}
\tablenotetext{\dag2}{The extinctions were corrected, using a typical $A_V$ of 3.5 \citep{Preibisch(2011)ApJS_194_10} and the extinction curve of ``Milky Way, $R_V=4.0$'' \citep{Weingartner(2001)ApJ_548_296,Draine(2003)ARA&A_41_241}.}
\tablenotetext{\dag3,\dag4}{These targets fall onto the overlapped regions (cf.~Fig.~\ref{fig-obs}). We measured their fluxes from each exposure, and averaged them.}
\end{deluxetable}

\clearpage
\begin{deluxetable}{ccccrr}
\tablewidth{0pt}
\tablecaption{Estimated Outflow Rate \label{tbl-out}}
\tablehead{
\colhead{Targets} & \colhead{Shape\tablenotemark{\dag1}} & \colhead{Solid} & \colhead{Outflow} & \colhead{Outflow Mass Loss} & \colhead{Outflow Mass Loss} \\
& & \colhead{Angle ($\Omega$)} & \colhead{Length ($L_{out}$)} & \colhead{Rate\tablenotemark{\dag2} (\Mout; [\astFe{II}])} & \colhead{Rate\tablenotemark{\dag3} (\Mout; \Ha)} \\
& & \colhead{($10^{-10}$ sr)} & \colhead{($''$)} & \colhead{($10^{-7}$ $M_{\odot}$ yr$^{-1}$)} & \colhead{($10^{-7}$ $M_{\odot}$ yr$^{-1}$)}
}
\startdata
                         IFO-1 &   K & \nodata & \nodata &  11.5$\pm$2.0 &     1.76 \\
    IFO-2\tablenotemark{\dag4} &   K & \nodata & \nodata &  18.0$\pm$1.7 &     0.21 \\
                         IFO-3 &   K & \nodata & \nodata &  20.1$\pm$1.9 &  \nodata \\
    IFO-4\tablenotemark{\dag5} &   L &     2.8 &     2.5 &   6.7$\pm$0.3 &     5.68 \\
                         IFO-5 &   L &     2.8 &     2.3 &   4.5$\pm$0.2 &  \nodata \\
                        IFO-6a &   L &     3.8 &     2.7 &  15.6$\pm$0.7 &  \nodata \\
                        IFO-6b &   K & \nodata & \nodata &  22.3$\pm$1.3 &  \nodata \\
                        IFO-6c &   K & \nodata & \nodata &  16.5$\pm$1.1 &  \nodata \\
                        IFO-7a &   L &     1.8 &     2.0 &   7.2$\pm$0.5 &     0.44 \\
                        IFO-7b &   K & \nodata & \nodata &  11.4$\pm$2.7 &  \nodata \\
                         IFO-8 &   K & \nodata & \nodata &  50.5$\pm$6.4 &  \nodata \\

\enddata
\tablenotetext{\dag1}{``K'' and ``L'' mean ``knotty'' and ``longish,'' respectively.}
\tablenotetext{\dag2}{The outflow rate is estimated from the dereddened [Fe II] flux, using two different methods according to the shape of IFOs. See the text for detail.}
\tablenotetext{\dag3}{The outflow rate is from \cite{Smith(2010)MNRAS_405_1153}, who estimated the rate from \Ha{} intensity assuming irradiated heating.}
\tablenotetext{\dag4,\dag5}{These targets fall onto the overlapped regions (cf.~Fig.~\ref{fig-obs}). We estimated the outflow mass loss rate, using the [\astFe{II}] flux from each exposure, and averaged them.}
\end{deluxetable}

\clearpage
\begin{deluxetable}{cccccccccccr}
\tablewidth{0pt}
\rotate
\tabletypesize{\scriptsize}
\tablecaption{Physical Parameters of the Relevant YSOs \label{tbl-phy}}
\tablehead{
\colhead{Catalog} & \colhead{$log\,L_{bol}$\tablenotemark{\dag2}} & \colhead{$log\,L_{acc}$\tablenotemark{\dag2}} & \colhead{$log\,M_{*}$\tablenotemark{\dag2}} & \colhead{$log\,M_{disk}$\tablenotemark{\dag2}} & \colhead{$log\,M_{env+amb}$\tablenotemark{\dag2}} & \colhead{$log\,t_*$\tablenotemark{\dag2}} & \colhead{$log\,\dot{M}_{acc,env}$\tablenotemark{\dag2}} & \colhead{$log\,\dot{M}_{acc,disk}$\tablenotemark{\dag2}} & \colhead{YSO} & \colhead{Targets\tablenotemark{\dag3}} \\
\colhead{Number\tablenotemark{\dag1}} & & & & & & & & & \colhead{Stage\tablenotemark{\dag2}} & \\
& \colhead{($L_{\odot}$)} & \colhead{($L_{\odot}$)} & \colhead{($M_{\odot}$)} & \colhead{($M_{\odot}$)} & \colhead{($M_{\odot}$)} & \colhead{(yr)} & \colhead{($M_{\odot}$ yr$^{-1}$)} & \colhead{($M_{\odot}$ yr$^{-1}$)}
}
\startdata
490 &  1.58$\pm$0.07 & $-$1.35$\pm$0.24 &  0.40$\pm$0.02 & $-$3.22$\pm$0.29 & $-$5.87$\pm$0.43 &  6.77$\pm$0.10 & $-$9.91$\pm$0.29 & $-$8.93$\pm$0.26 & II & IFO-2 \\
842 &  1.46$\pm$0.31 &  0.32$\pm$0.50 & $-$0.01$\pm$0.22 & $-$2.64$\pm$0.64 & $-$0.44$\pm$0.31 &  3.67$\pm$0.44 & $-$4.25$\pm$0.42 & $-$6.11$\pm$0.52 & 0/I & IFO-4 \\
984 &  1.22$\pm$0.14 & $-$0.27$\pm$0.47 &  0.01$\pm$0.15 & $-$2.21$\pm$0.32 &  0.26$\pm$1.55 &  4.33$\pm$0.49 & $-$3.75$\pm$1.70 & $-$6.86$\pm$0.59 & 0/I & IFO-7 \\

\enddata
\tablenotetext{\dag1}{The numbers are from \cite{Povich(2011)ApJS_194_14}.}
\tablenotetext{\dag2}{These characteristic quantities are obtained from the SED fitting \citep{Robitaille(2007)ApJS_169_328}, using the photometry data of \cite{Povich(2011)ApJS_194_14}. See text for detail.}
\tablenotetext{\dag3}{cf.~Figure \ref{fig-jet} and Table \ref{tbl-flux}.}
\end{deluxetable}

\clearpage
\begin{deluxetable}{cccc}
\tablewidth{0pt}
\tablecaption{[\astFe{II}] Detection Rate and the YSO stage \label{tbl-stg}}
\tablehead{
\colhead{Stage} & \colhead{[\astFe{II}]} & \colhead{YSOs within} & \colhead{Detection} \\
& \colhead{detected} & \colhead{the Field} & \colhead{Rate}
}
\startdata
0/I & 2 & 118 & 1.7\% \\
II & 1 & 160 & 0.6\% \\
III & 0 & 17 & 0.0\% \\
A & 0 & 264 & 0.0\% \\
Total & 8\tablenotemark{\dag} & 559 & 1.4\% \\
\enddata
\tablenotetext{\dag}{This includes five IFOs whose jet-driving YSOs are not definitely determined. See section \ref{ana-res-det} and Table \ref{tbl-phy}.}
\end{deluxetable}

\end{document}